\documentclass[12pt,preprint,useAMS,usenatbib]{emulateapj}
\usepackage{graphicx, txfonts}
\usepackage{url}
\usepackage{multirow}
\usepackage{natbib}

\usepackage[usenames,dvipsnames]{color}

\newcommand{\dhp}{(D/H)$_{\rm P}$}
\newcommand{\neff}{$N_{\rm eff}$}
\newcommand{\obhh}{$\Omega_{\rm B,0}\,h^{2}$}
\newcommand{\AlII}{Al\,\textsc{ii}}

\newcommand{\CII}{C\,\textsc{ii}}

\newcommand{\CIV}{C\,\textsc{iv}}

\newcommand{\DI}{\textrm{D}\,\textsc{i}}
\newcommand{\DII}{\textrm{D}\,\textsc{ii}}
\newcommand{\FeII}{Fe\,\textsc{ii}}

\newcommand{\HI}{\textrm{H}\,\textsc{i}}

\newcommand{\HII}{\textrm{H}\,\textsc{ii}}

\newcommand{\Lya}{Ly$\alpha$}

\newcommand{\NI}{N\,\textsc{i}}
\newcommand{\NII}{N\,\textsc{ii}}
\newcommand{\NIII}{N\,\textsc{iii}}

\newcommand{\OI}{O\,\textsc{i}}

\newcommand{\SII}{S\,\textsc{ii}}
\newcommand{\SiII}{Si\,\textsc{ii}}
\newcommand{\SiIII}{Si\,\textsc{iii}}
\newcommand{\SiIV}{Si\,\textsc{iv}}

\def\rahr{^{\rm h}}
\def\ramin{^{\rm m}}
\def\rasec{\!\!^{\rm s}}
\def\decdeg{^{\circ}}
\def\decmin{'}
\def\decsec{\!\!''}

\shorttitle{\textsc{A precise measure of primordial D/H}}
\shortauthors{\textsc{Cooke et al.}}

\begin{document}

\title{The primordial deuterium abundance of the most metal-poor damped Lyman-$\alpha$ system\altaffilmark{$\star$}}

\author{Ryan J. Cooke\altaffilmark{1,2,3},
Max Pettini\altaffilmark{4,5},
Kenneth M. Nollett\altaffilmark{6},
Regina Jorgenson\altaffilmark{7,8}}

\altaffiltext{$\star$}{Based on observations collected at the European Organisation for Astronomical Research 
in the Southern Hemisphere, Chile [VLT program ID: 093.A-0016(A)],
and at the W.M. Keck Observatory
which is operated as a scientific partnership among the California Institute of 
Technology, the University of California and the National Aeronautics and Space 
Administration. The Observatory was made possible by the generous financial
support of the W.M. Keck Foundation.}
\altaffiltext{1}{UCO/Lick Observatory, University of California Santa Cruz, Santa Cruz, CA 95064, USA}
\altaffiltext{2}{Kavli Institute for Particle Astrophysics and Cosmology (KIPAC), Stanford University, 452 Lomita Mall, Stanford, CA 94305, USA}
\altaffiltext{3}{Hubble Fellow;~~~~email: rcooke@ucolick.org}
\altaffiltext{4}{Institute of Astronomy, Madingley Road, Cambridge CB3 0HA, UK}
\altaffiltext{5}{Kavli Institute for Cosmology, Madingley Road, Cambridge CB3 0HA, UK}
\altaffiltext{6}{Department of Physics, San Diego State University, 5500 Campanile Drive, San Diego, CA 92182, USA}
\altaffiltext{7}{Willamette University, Physics Department, 900 State Street, Salem, OR 97301, USA}
\altaffiltext{8}{Maria Mitchell Observatory, 4 Vestal Street, Nantucket, MA 02554, USA}

\begin{abstract}
We report the discovery and analysis of the most metal-poor damped Lyman-$\alpha$
(DLA) system currently known, which also displays the Lyman series absorption lines
of neutral deuterium. The average [O/H] abundance of this system is [O/H]~$= -2.804\pm0.015$,
which includes an absorption component with [O/H]~$= -3.07\pm0.03$. Despite the unfortunate
blending of many weak \DI\ absorption lines, we report a precise measurement of the deuterium abundance of this system. Using the six highest quality and self-consistently analyzed
measures of D/H in DLAs, we report tentative
evidence for a subtle decrease of D/H with increasing metallicity.
This trend must be confirmed with future high precision D/H measurements
spanning a range of metallicity.
A weighted mean of these six independent measures provides our best estimate of the
primordial abundance of deuterium, $10^{5}\,({\rm D/H})_{\rm P} = 2.547\pm0.033$
($\log_{10} {\rm (D/H)_P} = -4.5940 \pm 0.0056$).
We perform a series of detailed Monte Carlo calculations of Big Bang nucleosynthesis (BBN)
that incorporate the latest determinations of several key nuclear
cross sections, and propagate their associated uncertainty.
Combining our measurement of (D/H)$_{\rm P}$ with these BBN
calculations yields an estimate of the cosmic baryon density,
$100\,\Omega_{\rm B,0}\,h^{2}({\rm BBN}) = 2.156\pm0.020$,
if we adopt the most recent theoretical determination of the
$d(p,\gamma)^3\mathrm{He}$ reaction rate.
This measure of \obhh\ differs by $\sim2.3\sigma$ from the Standard Model
value estimated from the \textit{Planck} observations of the cosmic microwave
background. Using instead a $d(p,\gamma)^3\mathrm{He}$ reaction rate that
is based on the best available experimental cross section data,
we estimate $100\,\Omega_{\rm B,0}\,h^{2}({\rm BBN}) = 2.260\pm0.034$,
which is in somewhat better agreement with the \textit{Planck} value.
Forthcoming measurements of the crucial $d(p,\gamma)^3\mathrm{He}$
cross section may shed further light on this discrepancy.
\end{abstract}


\section{Introduction}

\setcounter{footnote}{8}

Moments after the Big Bang, a brief period of nucleosynthesis created
the first elements and their isotopes
\citep{HoyTay64,Pee66,WagFowHoy67},
including hydrogen (H), deuterium (D), helium-3 ($^{3}$He), helium-4 ($^{4}$He),
and a small amount of lithium-7 ($^{7}$Li).
The creation of these elements, commonly referred to
as Big Bang nucleosynthesis (BBN), was concluded in
$\lesssim15$ minutes
and currently offers our earliest reliable probe of cosmology
and particle physics (for a review, see \citealt{Ste07,Ioc09,Ste12,Cyb15}).

The amount of each primordial nuclide that was made during
BBN depends most sensitively on the expansion rate of the
Universe and the number density ratio of baryons-to-photons.
Assuming the Standard Model of cosmology and particle physics,
the expansion rate of the Universe during BBN is driven by photons,
electrons, positrons, and 3 neutrino families. Furthermore, within the
framework of the Standard Model, the baryon-to-photon ratio at the
time of BBN (i.e. minutes after the Big Bang) is identical to the
baryon-to-photon ratio at recombination ($\sim400\,000$ years after
the Big Bang). Thus, the abundances of the primordial nuclides for
the Standard Model can be estimated from observations of the Cosmic
Microwave Background (CMB) radiation, which was recently recorded
with exquisite precision by the \textit{Planck} satellite \citep{Efs15}.
Using the \textit{Planck} CMB observations\footnote{The primordial
abundances listed here use the
TT+lowP+lensing measure of the baryon density,
$100\,\Omega_{\rm B,0}\,h^{2}({\rm CMB})=2.226\pm0.023$,
 (i.e. the second data column of Table~4 from \citealt{Efs15}).},
 the predicted Standard
Model abundances of the primordial elements are
(68 per cent confidence limits; see Section~\ref{sec:dh}):
\begin{eqnarray}
Y_{\rm P}&=&0.2471\pm0.0005\nonumber\\
10^{5}\,({\rm D/H})_{\rm P}&=&2.414\pm0.047\nonumber\\
10^{5}\,({\rm ^{3}He/H})_{\rm P}&=&1.110\pm0.022\nonumber\\
A(^{7}{\rm Li/H})_{\rm P}&=&2.745\pm0.021\nonumber
\end{eqnarray}
where $Y_{\rm P}$ is the fraction of baryons consisting
of $^{4}$He, $A(^{7}{\rm Li/H})_{\rm P}\equiv\log_{10}(^{7}{\rm Li/H})_{\rm P}+12$,
and D/H, $^{3}$He/H and $^{7}$Li/H are the number abundance ratios
of deuterium, helium-3 and lithium-7 relative to hydrogen, respectively.

To test the Standard Model, the above predictions are usually
compared to direct observational measurements of these
abundances in near-primordial environments. High precision
measures of the primordial $^{4}$He mass fraction are obtained
from low metallicity \HII\ regions in nearby star-forming galaxies.
Two analyses of the latest measurements, including an infrared
transition that was not previously used, find
$Y_{\rm P}~=~0.2551\pm0.0022$ \citep{IzoThuGus14}, and
$Y_{\rm P}~=~0.2449\pm0.0040$ \citep{AveOliSki15}.
These are mutually inconsistent, presumably due to
some underlying difference between the analysis methods.
The primordial $^{7}{\rm Li/H}$ ratio is deduced from the most
metal-poor stars in the halo of the Milky Way. The latest
determination \citep{Asp06,Aok09,Mel10,Sbo10,Spi15},
$A(^{7}{\rm Li})=2.199\pm0.086$, implies a $\gtrsim6\sigma$
deviation from the Standard Model value (see \citealt{Fie11} for a review).
The source of this discrepancy is currently unknown.
The abundance of $^{3}$He has only been measured in Milky Way
\HII\ regions \citep{BanRooBal02} and in solar system meteorite
samples \citep{BusBauWie00,BusBauWie01}. At this time, it is
unclear if these measures are representative of the primordial
value. However, there is a possibility that $^{3}$He might be
detected in emission from nearby, quiescent metal-poor
\HII\ regions with future, planned telescope facilities \citep{Coo15}.

The primordial abundance of deuterium, \dhp, can be estimated using
quasar absorption line systems \citep{Ada76}, which are clouds of gas
that absorb the light from an unrelated background quasar. In rare, quiescent
clouds of gas the $-82~{\rm km~s}^{-1}$ isotope shift of D relative to H can be
resolved, allowing a measurement of the column density ratio \DI/\HI. The
most reliable measures of \dhp\ come from near-pristine damped
Lyman-$\alpha$ systems (DLAs). As discussed in \citet{PetCoo12a} and \citet{Coo14},
metal-poor DLAs exhibit the following properties that facilitate a high
precision and reliable determination of the primordial deuterium abundance:
(1) The Lorentzian damped \Lya\ absorption line uniquely determines the
total column density of neutral H atoms along the line-of-sight.
(2) The array of weak, high order \DI\ absorption lines depend only
on the total column density of neutral D atoms along the line-of-sight.
Provided that these absorption lines fall on the linear regime of
the curve-of-growth, the derived $N$(\DI) should not depend on
the gas kinematics or the instrument resolution.
In addition, the assumption that D/H=\DI/\HI\ is justified in these
systems; the ionization correction is expected to be
$\lesssim0.1$~per~cent \citep{Sav02,CooPet16}. Furthermore,
galactic chemical evolution models suggest that most of the
deuterium atoms in these almost pristine systems are yet to be
cycled through many generations of stars; the correction for
astration (i.e. the processing of gas through stars) is therefore
negligible (see the comprehensive list of references
provided by \citealt{Cyb15,Dvo16}).

Using a sample of 5 quasar absorption line systems that satisfy a
set of strict criteria, \citet{Coo14} recently estimated that the primordial
abundance of deuterium is log$_{10}$\,\dhp~=~$-4.597\pm0.006$, or
expressed as a linear quantity, $10^{5}\,({\rm D/H})_{\rm P} = 2.53\pm0.04$.
These 5 systems exhibit a D/H plateau over at least a factor of $\sim10$ in metallicity,
and this plateau was found to be in good agreement with the expected value for the
cosmological model derived by \textit{Planck} assuming the Standard Model of particle
physics. In this paper, we build on this work and present a new determination of the
primordial abundance of deuterium obtained from the lowest metallicity DLA
currently known. In Section~\ref{sec:obs}, we present the details of our observations
and data reduction procedures. Our data analysis is almost identical to that
described in \citet{Coo14}, and we provide a summary of this procedure
in Section~\ref{sec:analysis}. In Section~\ref{sec:chemcomp}, we report
the chemical composition of this near-pristine DLA. In Section~\ref{sec:dh},
we present new calculations of BBN that incorporate the latest nuclear
cross sections, discuss the main results of our analysis, and highlight
the cosmological implications of our findings.
We summarize our conclusions in Section~\ref{sec:conc}.

\section{Observations and Data Reduction}
\label{sec:obs}

In this paper, we present high quality echelle observations of the
quasar J1358+0349 ($z_{\rm em}\simeq2.894$,
Right Ascension$=13\rahr58\ramin03.\rasec97$,
Declination$=+03\decdeg49\decmin36.\decsec0$),
which was discovered with a low resolution ($R\sim2000$) spectrum acquired by the
Sloan Digital Sky Survey (SDSS). This SDSS spectrum revealed strong \HI\ absorption
at a redshift $z_{\rm abs}=2.8528$ with no apparent absorption at the wavelengths of the
corresponding metal lines, indicating the presence of a very metal-poor DLA \citep{Pen10}.
\citet{Pen10} reobserved this quasar with the Echellette Spectrograph and Imager (ESI),
which is mounted on the Keck II telescope. These medium resolution observations
($R\sim5300$, corresponding to a velocity full width at half maximum
$v_{\rm FWHM}\simeq57~{\rm km~s}^{-1}$) confirmed that this DLA is
among the most metal-poor systems currently known, with an estimated
metallicity\footnote{Throughout this paper, we adopt the notation [X/Y] to
represent the relative number density of elements $X$ and $Y$
on a logarithmic and solar abundance scale. Explicitly,
[X/Y]~$=~\log_{10}(N({\rm X})/N({\rm Y}))-\log_{10}(n({\rm X})/n({\rm Y}))_{\odot}$.}
${\rm [Fe/H]}=-3.03\pm0.11$. We confirm the low metallicity with
the higher resolution data presented here;
we find ${\rm [Fe/H]}=-3.25\pm0.11$ (see Section~\ref{sec:chemcomp}),
assuming a solar abundance $\log_{10}({\rm Fe/H})_{\odot}=-4.53$
\citep{Asp09}.

Identifying DLAs where the \DI\ Lyman series absorption lines
are well-resolved from the much stronger \HI\ Lyman series is one
of the primary difficulties of finding DLAs where D/H can be measured.
The probability of resolving these features can be increased by finding
gas clouds with simple kinematics, which are more common at the lowest
metallicity \citep{Led06,Mur07,Pro08,Nel13,JorMurTho13,CooPetJor15};
in general, the most metal-poor systems exhibit simple and quiescent kinematics.
Given the low metallicity of the DLA towards J1358+0349, based on the ESI
spectra, we acquired two high-quality, high resolution spectra of this quasar
with the aim of measuring D/H. We describe these observations below.

\subsection{HIRES observations}

We observed J1358+0349 with the High Resolution Echelle Spectrometer
(HIRES; \citealt{Vog94}) on the Keck I telescope on 2013 May 6 in good
seeing conditions ($\sim0.7''$~FWHM) for a total of 21,000~s divided
equally into $7\times3000~{\rm s}$ exposures. We used the blue-sensitive
ultraviolet cross-disperser to maximize the efficiency near the DLA Lyman
limit. We used the C1 decker ($7.0''\times0.861''$), which provides a
nominal instrument resolution of $R~\simeq~48,000$
($v_{\rm FWHM}\simeq6.4~{\rm km~s}^{-1}$) for a uniformly illuminated slit.
By measuring the widths of $670$ ThAr wavelength calibration lines\footnote{Ideally, O$_{2}$
telluric absorption should be used to determine the instrument resolution,
since the broadening of these lines should closely represent the instrument
resolution of the quasar absorption spectrum; unlike the sky and ThAr lamp
emission lines, the quasar light does not uniformly illuminate the slit. However,
the telluric O$_{2}$ molecular absorption band near 6300\,\AA\ was too
weak to reliably measure the instrument FWHM.},
we determined the instrument resolution to be
$v_{\rm FWHM}=6.17\pm0.02~{\rm km~s}^{-1}$,
which is somewhat lower than the nominal value.
All frames were binned $2\times2$
during read-out. The science exposures were bracketed by a ThAr
wavelength calibration frame.
The final data cover the wavelength range 3480\,\AA\,--\,6344\,\AA,
with small gaps in the ranges 4397\,\AA\,--\,4418\,\AA\ and 5397\,\AA\,--\,5423\,\AA\
due to the gaps between the three HIRES detectors.

\subsection{UVES observations}

The HIRES data confirmed the very low metallicity of the DLA,
and revealed several resolved \DI\ absorption lines, suggesting
that this system would be ideal to estimate the primordial deuterium
abundance. To increase the signal-to-noise (S/N) ratio of the data,
we observed J1358+0349 for a total of 40,384\,s with the
Very Large Telescope (VLT) Ultraviolet and Visual Echelle
Spectrograph (UVES; \citealt{Dek00}) in service mode.\footnote{Our
observations were carried out on
2014 March 28 ($3\times3495\,{\rm s}$),
2014 May 27 ($3\times3495\,{\rm s}$),
2014 March 24 ($4\times3495\,{\rm s}$),
2014 April 30 ($1\times3495\,{\rm s}$, $1\times1939\,{\rm s}$).}
We used dichroic 1, with the HER\_5 filter in the blue arm, and
the SHP700 filter in the red arm. The echelle grating in the blue
arm provided a central wavelength of 3900\,\AA, whilst the grating
in the red arm had a central wavelength of 5640\,\AA. The UVES
data cover the wavelength range 3450\,\AA\,--\,6648\,\AA, with small
gaps in the ranges 4530\,\AA\,--\,4622\,\AA\ and 5601\,\AA\,--\,5675\,\AA.
All exposures were binned $2\times2$ at the time of read-out.
We used the $0.9''$ slit to match closely the nominal resolution
provided by the HIRES observations (the nominal UVES values
are $R~\simeq~46,000$, $v_{\rm FWHM}~\simeq~6.5~{\rm km~s}^{-1}$).
By fitting $268$ ThAr emission lines, we derived an instrumental
resolution of $v_{\rm FWHM}=6.39\pm0.04~{\rm km~s}^{-1}$
for a uniformly illuminated slit with our setup. Our value is in good
agreement with the nominal UVES instrument resolution.\footnote{We note that
the telluric O$_{2}$ molecular absorption band near 6300\,\AA\ was
too weak, like the HIRES data, to reliably measure the instrument
FWHM.}

\subsection{Data Reduction}

The HIRES and UVES data described above provide complete
wavelength coverage from the DLA Lyman limit ($\sim$3520\,\AA) to
6648\,\AA\ (1725\,\AA\ in the rest-frame of the DLA). The data were
reduced with the HIRESRedux and UVESRedux\footnote{These
reduction packages can be obtained from\\
http://www.ucolick.org/$\sim$xavier/HIRedux/index.html}
software packages, maintained by J.~X.~Prochaska
(for a description of the reduction algorithms, see \citealt{BerBurPro15}).
The standard reduction steps were followed. First, the bias level
was subtracted from all frames using the overscan region. The
pixel-to-pixel variations were then removed using an archived
image, where the detector was uniformly illuminated. The orders
were defined using a quartz lamp with an identical slit and setup
as the science exposures. A ThAr lamp was used to model the
regions of constant wavelength across the detector (e.g. \citealt{Kel03}).
Using this model, the sky background was subtracted from the science
exposure. The spectrum of the quasar was extracted using an
optimal extraction algorithm, and mapped to a vacuum, heliocentric
wavelength scale with reference to the ThAr exposure.

Each echelle order was corrected for the echelle blaze
function, resampled onto a $2.5~{\rm km~s}^{-1}$
pixel scale, and combined using the UVES\_\textsc{popler}
software.\footnote{UVES\_\textsc{popler} can be downloaded from\\
http://astronomy.swin.edu.au/$\sim$mmurphy/UVES\_popler/}
Since the HIRES and UVES data were acquired with slightly
different instrument resolutions, we separately combined the
UVES and HIRES data. Deviant pixels and ghosts were
manually removed, and an initial estimate of the quasar
continuum was applied. The data were flux calibrated using
the SDSS discovery spectrum as a reference. Specifically,
the UVES and HIRES data were convolved with the SDSS
instrument resolution, and then resampled onto the wavelength
scale of the SDSS spectrum to determine the sensitivity
function. The sensitivity function was then applied to the
non-convolved UVES and HIRES data, with an extrapolation
to blue wavelengths where the SDSS spectrum does not
extend. The final HIRES spectrum has a S/N near the DLA
\Lya\ absorption line of ${\rm S/N}\simeq30$, and a ${\rm S/N}\simeq16$ near
the Lyman limit of the DLA. The equivalent values for UVES
are ${\rm S/N}\simeq40$ and ${\rm S/N}\simeq11$, respectively.

\section{Analysis Method}
\label{sec:analysis}

\begin{figure*}
  \centering
 {\includegraphics[angle=0,width=140mm]{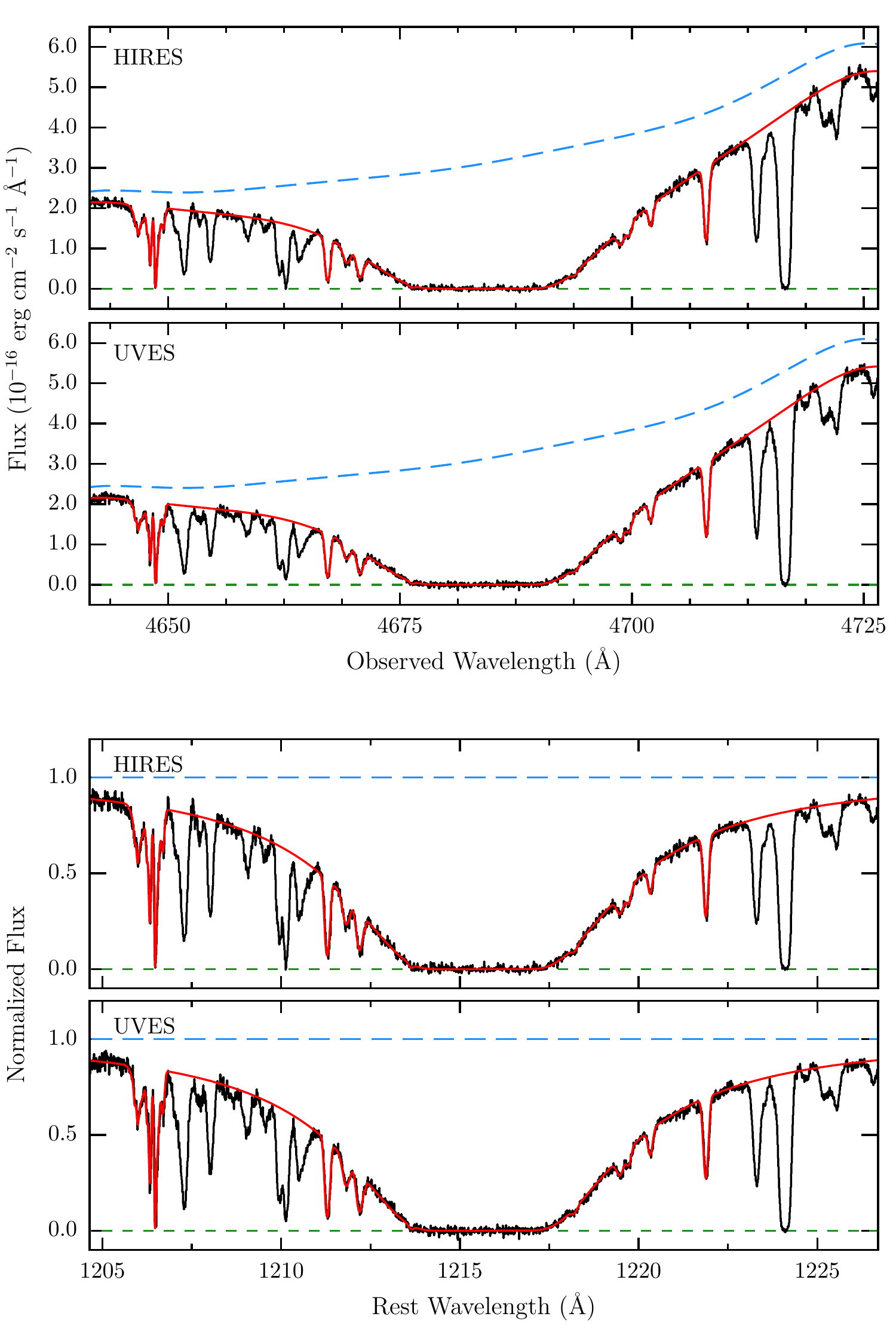}}\\
  \caption{
\textit{Top panels}: The flux calibrated \HI\ \Lya\ absorption profile (black histogram)
is shown for the DLA at $z_{\rm abs}=2.853054$ towards the quasar J1358$+$0349.
The best-fitting quasar continuum model (blue long-dashed curves) and the best-fitting
absorption profile (red line) are overlaid.
The green dashed line indicates the fitted zero-level of the data.
The spectrograph used to take the data is indicated in the upper left corner of each panel.
\textit{Bottom panels}: Same as the top panels, but with the quasar continuum normalized,
and the data are plotted in the rest-frame of the DLA.
The absorption feature that is fit near a rest wavelength of 1206.5 is a combination of the
\SiIII\ absorption from the DLA and an unrelated blend.
  }
  \label{fig:lya}
\end{figure*}

Our analysis method is identical to that outlined by \citet{Coo14}.
In this section, we summarize the main aspects of this
procedure. We use the Absorption LIne Software (\textsc{alis})
package to provide a simultaneous fit to the emission spectrum
of the quasar and the absorption lines of the
DLA.\footnote{\textsc{alis} is available for download at the following
website:\\https://github.com/rcooke-ast/ALIS}
\textsc{alis} uses a chi-squared minimization procedure to deduce
the model parameter values that best fit the data, weighted by the
quasar error spectrum.

Our line fitting procedure was applied at the
same time to both the UVES and HIRES data, to find the model that
fitted \textit{both} sets of data best.
We simultaneously fit the \HI\ and \DI\
Lyman series absorption lines, all of the significantly detected
metal absorption lines, the zero-levels of the HIRES and UVES
data, the continuum in the neighborhood of all absorption lines,
the relative velocity offset between the HIRES and UVES data,
and the instrument resolution of both datasets. The continuum
is approximated by a low order Legendre polynomial (typically
of order $\lesssim4$, except near \Lya\ where we use a polynomial
of order 8). To allow for relative differences in the
quasar continuum between the HIRES and UVES data, we
apply a constant or linear scaling to the HIRES data, and the
parameters of this scaling are allowed to vary during the
minimization procedure.

The portion of the \Lya\ absorption profile where the optical depth is
$\tau\gtrsim1$ provides most of the power to determine the total
\HI\ column density; when the quasar flux recovers to $\gtrsim50$
per cent of the continuum, the absorption profile flattens and becomes
increasingly sensitive to the continuum level rather than the \HI\ absorption. We therefore
fit every pixel in the core of the \Lya\ absorption until the Lorentzian wings of
the profile are 50 per cent of the continuum (i.e. $\tau\gtrsim0.7$; in this case, all pixels within
$\pm1300~{\rm km~s}^{-1}$). During the analysis, we fit all of the
contaminating absorption features within this velocity window instead
of masking the affected pixels. Outside this velocity window, we include pixels in the
fit that we deem are free of contamination. The best-fitting model of the
\Lya\ absorption feature is overlaid on the HIRES and UVES data in
Figure~\ref{fig:lya}.

\begin{table*}
\centering
\begin{minipage}[c]{0.6\textwidth}
    \caption{\textsc{Best fitting model parameters for the DLA at $z_{\rm abs}=2.853054$ towards the QSO J1358$+$0349}}
    \hspace{-0.6cm}\begin{tabular}{@{}crrccccc}
    \hline
   \multicolumn{1}{c}{Comp.}
& \multicolumn{1}{c}{$z_{\rm abs}$} 
& \multicolumn{1}{c}{$b_{\rm turb}$} 
& \multicolumn{1}{c}{$\log N$\/(H\,{\sc i})}
& \multicolumn{1}{c}{$\log {\rm (D\,\textsc{i}/H\,\textsc{i})}$}
& \multicolumn{1}{c}{$\log N$\/(N\,{\sc i})}
& \multicolumn{1}{c}{$\log N$\/(N\,{\sc ii})}
& \multicolumn{1}{c}{$\log N$\/(N\,{\sc iii})}\\
    \multicolumn{1}{c}{}
& \multicolumn{1}{c}{}
& \multicolumn{1}{c}{(km~s$^{-1}$)}
& \multicolumn{1}{c}{(cm$^{-2}$)}
& \multicolumn{1}{c}{}
& \multicolumn{1}{c}{(cm$^{-2}$)}
& \multicolumn{1}{c}{(cm$^{-2}$)}
& \multicolumn{1}{c}{(cm$^{-2}$)}\\
  \hline
1  & $2.852874$                     & $4.7$             & $20.16$         & $-4.582^{\rm a}$         &  $12.61$          &  \ldots$^{\rm b}$&  \ldots$^{\rm b}$   \\
    &  $\pm 0.000002$            &$\pm 0.2$       & $\pm 0.02$    & $\pm 0.012$                &  $\pm  0.10$  &                        &    \\
2  & $2.853004$                  & $3.9$             & \ldots$^{\rm b}$ & \ldots$^{\rm b}$         &  \ldots$^{\rm b}$ & $13.25^{\rm c}$        &  $13.32^{\rm c}$  \\
    &  $\pm 0.000003$                &$\pm 0.3$     &                          &                                        &                          &  $\pm0.04$  &  $\pm0.06$  \\
3  & $2.853054$                  & $2.5$             & $20.27$          & $-4.582^{\rm a}$         &  $12.23$        &  \ldots$^{\rm b}$&  $13.33$  \\
    &  $\pm 0.000003$        &$\pm 0.5$      & $\pm 0.02$    & $\pm 0.012$                &  $\pm0.24$    &                        &  $\pm0.06$  \\
4  & $2.85372$                      & $14.2$          & $18.23$          & $-4.582^{\rm a}$         &  \ldots$^{\rm b}$ &  \ldots$^{\rm b}$ &  $12.60$   \\
    &  $\pm 0.00001$            &$\pm 1.4$      & $\pm 0.07$     & $\pm 0.012$                &                        &                         &  $\pm0.13$    \\
Total  & \ldots                  & \ldots          & $20.524$          & $-4.582^{\rm a}$         &  $12.77$             &  $13.25$             &  $13.67$   \\
    &         &                                                   & $\pm 0.006$     & $\pm 0.012$                &    $\pm0.11$     &       $\pm 0.04$   &  $\pm0.02$    \\
  \hline
    \end{tabular}

    \smallskip\smallskip\smallskip

    \begin{tabular}{@{}ccccccc}
    \hline
   \multicolumn{1}{c}{Comp.}
& \multicolumn{1}{c}{$\log N$\/(O\,{\sc i})}
& \multicolumn{1}{c}{$\log N$\/(Al\,{\sc ii})}
& \multicolumn{1}{c}{$\log N$\/(Si\,{\sc ii})}
& \multicolumn{1}{c}{$\log N$\/(Si\,{\sc iii})}
& \multicolumn{1}{c}{$\log N$\/(S\,{\sc ii})}
& \multicolumn{1}{c}{$\log N$\/(Fe\,{\sc ii})}\\
    \multicolumn{1}{c}{}
& \multicolumn{1}{c}{(cm$^{-2}$)}
& \multicolumn{1}{c}{(cm$^{-2}$)}
& \multicolumn{1}{c}{(cm$^{-2}$)}
& \multicolumn{1}{c}{(cm$^{-2}$)}
& \multicolumn{1}{c}{(cm$^{-2}$)}
& \multicolumn{1}{c}{(cm$^{-2}$)}\\
  \hline
1  & $14.23$              & $11.27$               & $12.78$              &  \ldots$^{\rm b}$   &  \ldots$^{\rm b}$   & $12.31$ \\
    & $\pm 0.02$             & $\pm 0.07$            & $\pm 0.03$            &                                &                               & $\pm 0.18$ \\
2  & \ldots$^{\rm b}$   & $11.93$               & $12.94$               &  $12.89$               & $13.02$               & $12.51$ \\
    &                                & $\pm0.02$             &  $\pm0.04$             &  $\pm0.12$            &  $\pm0.10$            &  $\pm0.11$  \\
3  & $13.89$               & \ldots$^{\rm b}$   & $12.57$                &  $12.65$             & \ldots$^{\rm b}$  & \ldots$^{\rm b}$ \\
    & $\pm 0.02$          &                                & $\pm0.08$               &  $\pm0.10$          &      &   \\
4  & $12.86$               & \ldots$^{\rm b}$   & \ldots$^{\rm b}$   &  $12.19$            & \ldots$^{\rm b}$  & \ldots$^{\rm b}$ \\
    & $\pm 0.12$         &                                &                                 &  $\pm0.04$         &       &   \\
Total  & $14.41$         & $12.01$             & $13.27$                &  $13.14$      & $13.02$  & $12.74$ \\
    &     $\pm0.01$        &    $\pm0.02$      & $\pm0.01$            &  $\pm0.07$  & $\pm0.10$  & $\pm0.10$  \\
  \hline
    \end{tabular}

    \smallskip

$^{\rm a}${Forced to be the same for all components.}\\
\hspace{0.5cm}$^{\rm b}${Absorption is undetected for this ion in this component.}\\
\hspace{0.5cm}$^{\rm c}${Since the \NII\ and \NIII\ absorption lines arise from more highly ionized gas, we tie their total Doppler
parameter, and allow it to vary independently of the Doppler parameter of the other absorption lines at the redshift of this
component. The total Doppler parameter for these higher stages of N ionization is $b=10.5\pm0.9\,{\rm km~s}^{-1}$.}\\
    \label{tab:compstruct}
\end{minipage}
\end{table*}

\begin{figure*}[Ht]
  \centering
 {\includegraphics[angle=0,width=160mm]{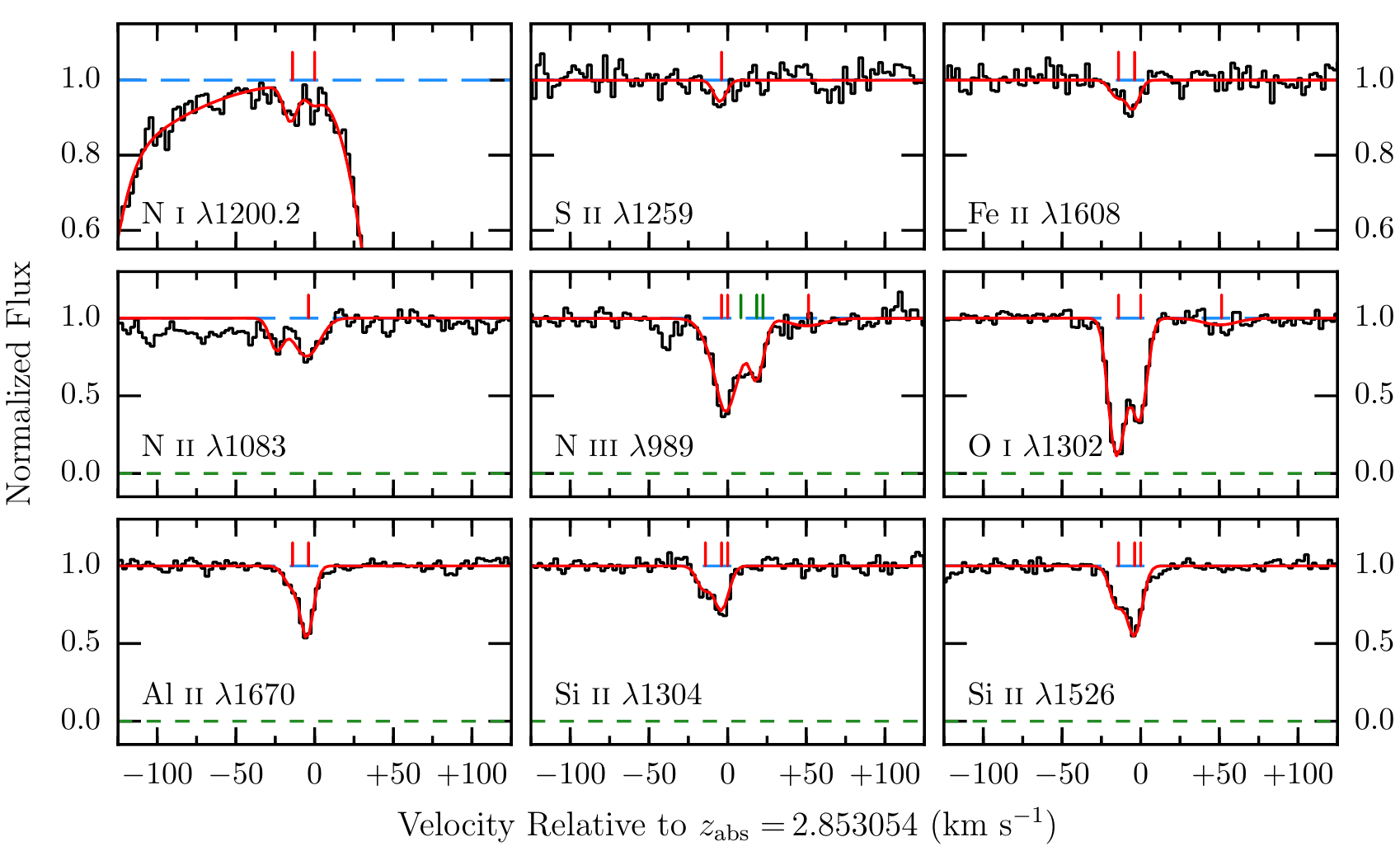}}\\
  \caption{
A selection of the metal absorption lines associated with the DLA
at $z_{\rm abs}=2.853054$ towards J1358$+$0349 that are used
in our analysis. The best-fitting model (red line) is derived from a
simultaneous fit to both the UVES and HIRES data. However,
in these panels we only show the data (black histogram) and
corresponding model for the dataset with the higher S/N near the
absorption line. In all panels, the best-fitting zero-level of
the data (short green dashed line) has been removed, and the
continuum has been normalized (long blue dashed line).
Note that we have used a different y-axis scale for the top row
of panels to emphasize the weakest absorption features.
The red tick marks above the spectrum correspond to the locations
of the absorption components of the annotated ion
(see Table~\ref{tab:compstruct}). The green tick marks in the
\NIII\,$\lambda989$ panel are for a blend with \SiII\,$\lambda989$,
the latter of which is largely determined from the multitude of other
\SiII\ absorption lines. The absorption at $-25~{\rm km~s}^{-1}$ in
the \NII\,$\lambda1083$ panel is assumed to be an unrelated blend.}
  \label{fig:metals}
\end{figure*}

Our spectrum includes 16 metal absorption lines from
the elements C, N, O, Al, Si, S and Fe in a range of
ionization stages, (\CII, \CIV, \NI, \NII, \NIII, \OI, \AlII, \SiII,
\SiIII, \SiIV, \SII\ and \FeII). The component structure of
our absorption model (see Table~\ref{tab:compstruct})
is set by the unblended, narrow metal absorption lines that
are the dominant ionization stage in neutral gas. The metal
absorption lines that are used in our analysis are presented in
Figure~\ref{fig:metals}. We find that the neutral \NI\
and \OI\ lines, which accurately trace the \DI\ bearing gas
\citep[][see also, \citealt{FieSte71,SteWerGel71}]{CooPet16},
are reproduced with just two principal absorption
components. The strong \OI\,$\lambda1302$ absorption line
also exhibits a much weaker absorption feature, comprising $\sim3$
per cent of the total \OI\ column density, and is redshifted
by $v\simeq+50\,{\rm km~s}^{-1}$ relative to the two main components;
this feature is also detected in the strong \CII\,$\lambda1334$ and
\SiII\,$\lambda1260$ absorption lines (not shown).
The first and higher ions, such as \NII, \AlII, \SiII, and \SII,
require an additional absorption component that is slightly blueshifted
by $v\simeq-4\,{\rm km~s}^{-1}$ relative to the systemic redshift
$z_{\rm abs}=2.853054$, and is presumably due to ionized gas.

We explicitly fit to the \DI/\HI\ ratio by requiring that all \DI\ absorption components
(i.e. components 1, 3, and 4 in Table~\ref{tab:compstruct}) have the same D/H ratio.
Note that the subdominant \DI\ absorption component (component 4, located at
$+50~{\rm km~s}^{-1}$ relative to the systemic redshift of the DLA) is not resolved
from the \HI\ absorption; the absorption properties of this component are only
determined by the \HI\ and \OI\ absorption. The initial starting value
of the logarithmic \DI/\HI\ ratio was randomly generated on the interval $(-4.8,-4.4)$.
We assume that the absorption lines of all species are represented
by a Voigt profile, comprising contributions from both turbulent
and thermal broadening. The standard assumption is that all gas
constituents in a given absorption component will share a common
turbulent Doppler parameter and a constant kinetic temperature.
As we discuss in \citet{Coo14}, at the current level of precision, a
Voigt profile that is broadened according to the above description
is probably insufficient to accurately model the
\HI, \DI, and metal absorption lines simultaneously; in reality,
there is a distribution of turbulence and temperature along
the line of sight. To circumvent
this model limitation, we tie the component redshifts and turbulent
Doppler parameters of all ions, and allow the thermal broadening
to be specified separately for the \DI\ and \HI\ absorption.
This prescription allows the kinematics of the \HI, \DI, and metal
absorption lines to be deduced almost independently.
We also stress that, as discussed in \citet{Coo14}, weak
unblended \DI\ absorption lines do not depend on the form of the Voigt profile;
the equivalent widths of weak \DI\ absorption lines uniquely
determine the \DI\ column density. Similarly, the absorption profile
of the \HI\ damped \Lya\ absorption line is independent of the
turbulence and kinetic temperature used for the Voigt profile fitting.

\begin{figure*}
  \centering
 {\includegraphics[angle=0,width=160mm]{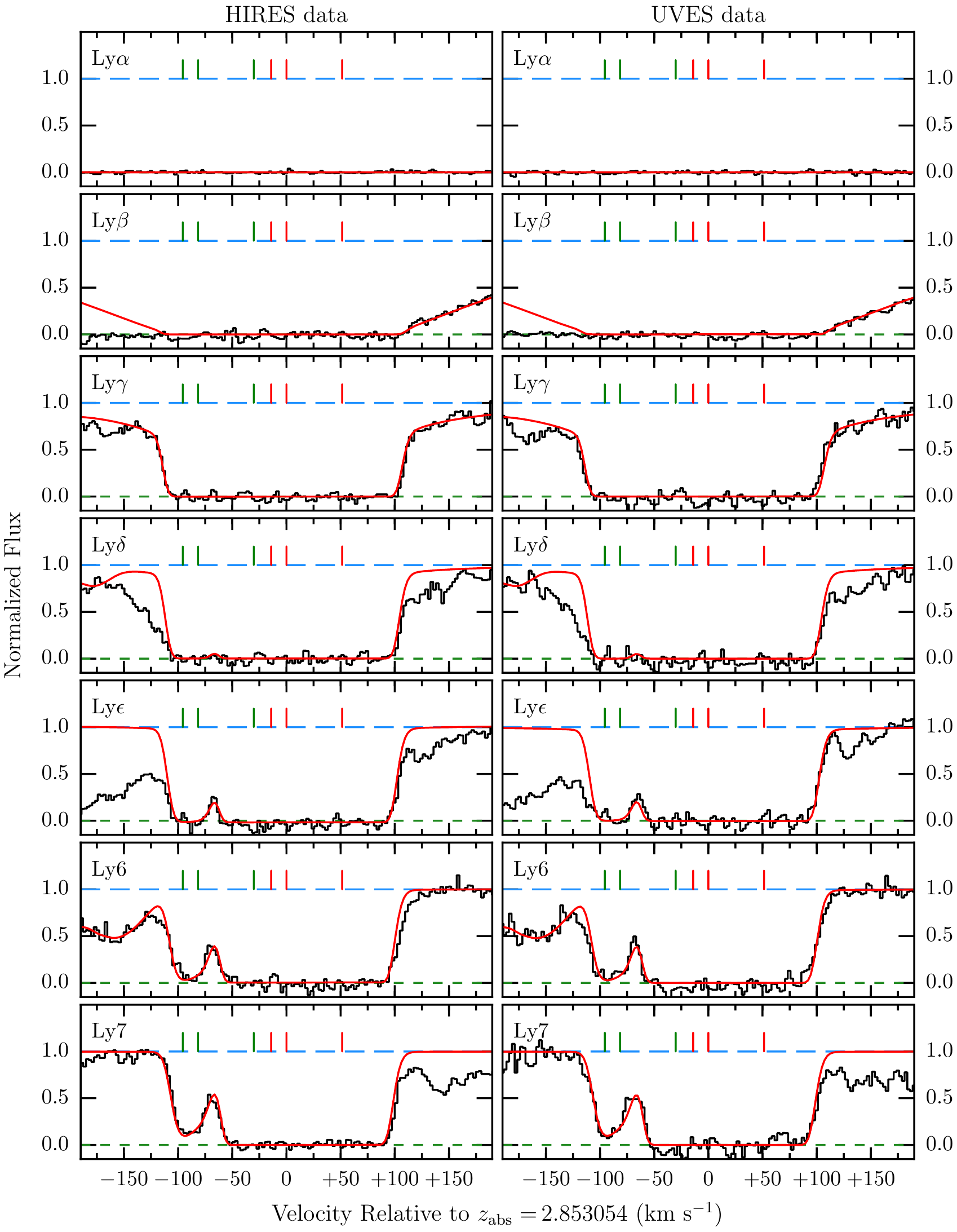}}\\
  \caption{
The black histogram shows our HIRES data (left panels) and UVES data (right panels),
covering the \HI\ and \DI\ Lyman series absorption lines from \Lya--Ly7 (top to bottom panels,
respectively). Our best-fitting model is overlaid with the solid red line. The plotted data have been
corrected for the best-fitting zero-level (short green dashed line), and are normalized by the
best-fitting continuum model (long blue dashed line). Tick marks above the spectrum
indicate the absorption components for \HI\ (red ticks), and \DI\ (green ticks).
  }
  \label{fig:lyseriesa}
\end{figure*}

\begin{figure*}
  \centering
 {\includegraphics[angle=0,width=160mm]{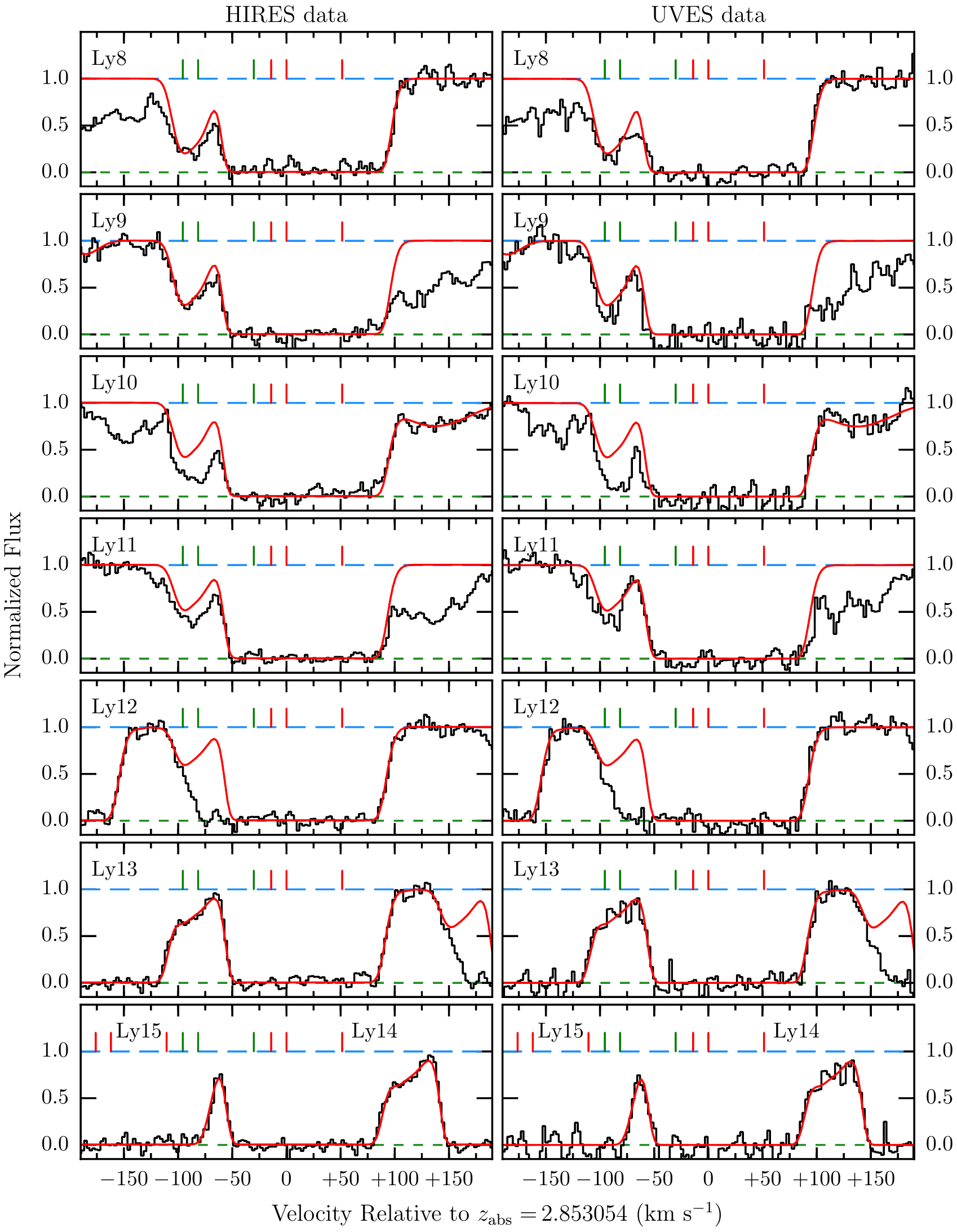}}\\
  \caption{
Same as Fig.~\ref{fig:lyseriesa}, for the \HI\ and \DI\ transitions Ly8--Ly15. Note that the
leftmost set of red tick marks in the bottom panels indicate the \HI\ Ly15 absorption
components, while the central red tick marks in these panels indicate \HI\ Ly14 absorption.
  }
  \label{fig:lyseriesb}
\end{figure*}

Our HIRES and UVES data of the Lyman series absorption
lines, together with the best-fitting model, are presented in
Figures~\ref{fig:lyseriesa} and \ref{fig:lyseriesb}.
In our analysis, we only use the \HI\ absorption lines that exhibit
either a clean blue or clean red wing. Similarly, we only consider
the \DI\ absorption lines that are free of unrelated contaminating
absorption. These include \DI\ Ly6, Ly7, Ly9, and Ly13; of these,
only Ly9 and Ly13 are weak, unsaturated absorption lines. We
also note that \DI\ Ly13 is
barely resolved from the \HI\ Ly14 absorption (see bottom
panels of Fig.~\ref{fig:lyseriesb}). Since the \HI\ Ly14 absorption
is well constrained by the host of other \HI\ Lyman series lines, we deem
the \DI\ equivalent width of Ly13 (particularly from the HIRES
data) to be well-determined. However, the DLA system that we
analyze here is certainly less ideal for the determination of D/H than our previously reported
cases in \citet{PetCoo12a} and \citet{Coo14}. In this new system,
many of the weak \DI\ absorption lines are blended with unrelated
absorption features (presumably contamination from low redshift
\Lya\ absorption),\footnote{This is one of the unpredictable, and
inherent difficulties associated with measuring the D/H ratio in
$z\sim3$ quasar absorption line systems.} resulting in fewer
unsaturated \DI\ lines. However, we are still able to constrain the value
of the D/H ratio within tight limits, thanks to the high S/N of our data near
the Lyman limit of the DLA, and the relatively well-determined value of
the \HI\ column density.

Initially, the instrumental FWHM was allowed to vary freely,
with no prior (as implemented in \citealt{Coo14}).
In this case, the fitted value of the instrumental FWHM was
larger than that allowed by the widths of the ThAr arc lines
(see Section~\ref{sec:obs}), implying that the DLA absorption
lines are too structured to permit a reliable estimate of the FWHM.
Thereafter, we fixed the instrumental FWHM to be equal to the
widths of the ThAr emission lines.

Finally, the relative velocity shift between the HIRES and UVES
data is determined during the $\chi^{2}$-minimization process,
with a best fitting value of $0.20\pm0.12~{\rm km~s}^{-1}$. We also
fit a wavelength independent correction to the zero-level of
each spectrum. This approximation also accounts for the
fraction of the quasar light that is not covered by the DLA
absorption. The best-fit values\footnote{This parameter
is largely driven by the trough of the \Lya\ absorption.} for
the zero-level are $0.016\pm0.003$ (HIRES) and
$0.003\pm0.003$ (UVES).

Our analysis was performed blindly, such that the $N$(\DI)/$N$(\HI)
ratio was only revealed after our profile analysis had been
finalized, and the minimum $\chi^{2}$ had been reached;
no changes were made to the data reduction or analysis
after the results were unblinded.
We then performed $2000$ Monte Carlo simulations to ensure that the
global minimum $\chi^{2}$ had been found. Each Monte Carlo simulation
was initialized with the best-fitting model parameters, perturbed
by twice the covariance matrix of the parameter values. The final
parameter values listed in Table~\ref{tab:compstruct} correspond
to the model with $\chi^{2}/{\rm dof}=7770/8678$, that provides
the global minimum chi-squared\footnote{As discussed in \citet{Coo14}, the $\chi^{2}$
value reported here should not be used for a statistical analysis,
since: (1) correlations between pixels are not accounted for; and
(2) the selected wavelength regions used for fitting tend to be those
with smaller statistical fluctuations.}.


\section{DLA Chemical Composition}
\label{sec:chemcomp}

The chemistry of the DLA at $z_{\rm abs}=2.853054$ towards J1358$+$0349 is
remarkable for several reasons. On the basis of six \OI\ lines, we
determine the average metallicity of the DLA to be [O/H]~$= -2.804\pm0.015$,
assuming a solar O abundance of $\log\,({\rm O/H})_{\odot}=-3.31$ \citep{Asp09}.
This cloud is therefore the most pristine DLA currently known (see \citealt{Coo11b}).
Furthermore, under our assumption that \DI/\HI\ is constant between the two main components,
the O abundance of the strongest \HI\ absorption (component 3 in Table~\ref{tab:compstruct})
is [O/H]~$= -3.07\pm0.03$.

We list the absolute and relative element abundances of this DLA in Table~\ref{tab:abund}.
Due to the presence of ionized gas (see Section~\ref{sec:analysis}), we quote an upper limit
on the abundance of Al, Si, S, and Fe; the first ions of these elements are the dominant stage of ionization in neutral (\HI) gas, but are also present in ionized (\HII) gas.
We also note that the [N/O] ratio
is well-determined in this DLA, since both \NI\ and \OI\ trace the \HI\ bearing gas due to
charge transfer reactions \citep{FieSte71,SteWerGel71}. Our value of [N/O] is consistent
with, or slightly lower than, the primary N/O plateau \citep{IzoThu99,Cen03,vZHay06,Pet08,PetLedSri08,PetCoo12b,Zaf14}.

In the final column of Table~\ref{tab:abund},
we also list the relative element abundances of component 1 ($z_{\rm abs}=2.852874$);
this absorption component probably arises from predominantly neutral gas, since
the higher stages of ionization are not detected in this component (see Table~\ref{tab:compstruct}).
Therefore, if the metals are well-mixed in this near-pristine DLA,\footnote{Note that chemical
variations have not been observed in other low metallicity DLAs \citep{Pro03,Coo11b}.} then
component 1 should reflect the chemistry of this system.
Relative to a typical metal-poor DLA \citep{Coo11b}, we find that this absorption component is somewhat
enhanced in oxygen relative to Al, Si, and Fe. It is not unexpected that the lighter elements,
such as C and O, exhibit an enhancement relative to the heavier elements (e.g. Fe) in
the lowest metallicity DLAs \citep{Coo11a,CooMad14}; this could be a signature of the
(washed out?) chemical enrichment from the first generation of stars \citep[e.g.][]{UmeNom03}.

\begin{table}
\begin{center}
    \caption{\textsc{Chemical composition of the DLA at $z_{\rm abs}=2.853054$ towards J1358$+$0349}}
    \hspace{-0.6cm}\begin{tabular}{@{}lcccc}
    \hline
   \multicolumn{1}{c}{X}
& \multicolumn{1}{c}{log\,$\epsilon$(X)$_{\odot}\,^{a,b}$}
& \multicolumn{1}{c}{[X/H]$^{c}$}
& \multicolumn{1}{c}{[X/O]$^{c}$}
& \multicolumn{1}{c}{[X/O]$_{1}\,^{d}$}\\
  \hline
N  &  $7.83$  &  $-3.58\pm0.11$        &  $-0.78\pm0.11$  &  $-0.76\pm0.10$  \\
O  &  $8.69$  &  $-2.804\pm0.015$   &  \ldots                     &  \ldots  \\
Al  &  $6.44$  &  $<-2.95$       &  $<-0.15$              &  $-0.71\pm0.07$  \\
Si  &  $7.51$  &  $<-2.764$  &  $<+0.04$             &  $-0.27\pm0.04$  \\
S   &  $7.14$  &  $<-2.64$       &  $<+0.16$             &  \ldots  \\
Fe &  $7.47$  &  $<-3.25$       &  $<-0.45$              &  $-0.70\pm0.18$  \\
  \hline
    \end{tabular}
    \label{tab:abund}
\end{center}

$^{\rm a}${log\,$\epsilon$(X) = 12 + log\,$N({\rm X})/N({\rm H}).$}\\
\hspace{0.5cm}$^{\rm b}${\citet{Asp09}.}\\
\hspace{0.5cm}$^{\rm c}${Limits are quoted for the first ions due to the presence of ionized gas.}\\
\hspace{0.5cm}$^{\rm d}${The final column lists the element abundance ratios of the mostly neutral absorption component at $z_{\rm abs}=2.852874$ (i.e. component number 1).}\\
\end{table}


\section{The Deuterium Abundance}
\label{sec:dh}

The near-pristine gas in the DLA reported here is a highly suitable environment for measuring the primordial abundance
of deuterium (see also \citealt{FumOmePro11} for the most metal-poor Lyman Limit system).
However, as discussed in Section~\ref{sec:analysis}, the structure of the absorption lines
and the unfortunate level of unrelated contamination limit the \textit{accuracy} with which
the deuterium abundance can be measured in this system.
The measured value of \DI/\HI\ in this DLA,
expressed as a logarithmic and linear quantity, is:
\begin{equation}
\log_{10}\,({\rm D\,\textsc{i}/H\,\textsc{i}}) = -4.582\pm0.012
\end{equation}
\begin{equation}
10^{5}~{\rm D\,\textsc{i}/H\,\textsc{i}} = 2.62\pm0.07
\end{equation}
which is consistent with the inverse variance weighted mean value of the five other
high precision measurements reported by \citet{Coo14},
$10^{5}~{\rm D\,\textsc{i}/H\,\textsc{i}} = 2.53\pm0.04$. The \DI/\HI\ measurement
precision obtained from this new DLA is comparable to the systems analyzed by
\citet{Coo14}, reflecting the high S/N of our data and the well-determined value
of the total \HI\ column density.

Despite the very low metallicity of this system, we also detect weak absorption
from \NI\ and \NII, resulting in an ion ratio $\log({\rm N\,\textsc{ii}/N\,\textsc{i}})=0.48\pm0.12$.
As recently highlighted by \citet{CooPet16}, charge transfer ensures that this ion ratio is
sensitive to the relative ionization of deuterium and hydrogen in DLAs, and can be used
to assess if an ionization correction must be applied to the measured \DI/\HI\ ratio to recover
the true D/H abundance. Using Equation~28 from \citet{CooPet16}, we estimate that the D/H
ionization correction for this system is:
\begin{equation}
{\rm IC(D/H)}\equiv\log_{10}{\rm (D/H)} - \log_{10}\,N({\rm D\,\textsc{i}})/N({\rm H\,\textsc{i}})
\end{equation}
\begin{equation}
{\rm IC(D/H)}=(-4.9\pm1.0)\times10^{-4}
\end{equation}
which includes a 6 per cent uncertainty in the ionization correction relation,
as recommended by \citet{CooPet16}. Since this correction is a factor of
$\sim25$ below the precision of this single measurement, we do not
apply this correction to our results.

\subsection{Metallicity Evolution}

\begin{figure*}
  \centering
 {\includegraphics[angle=0,width=85mm]{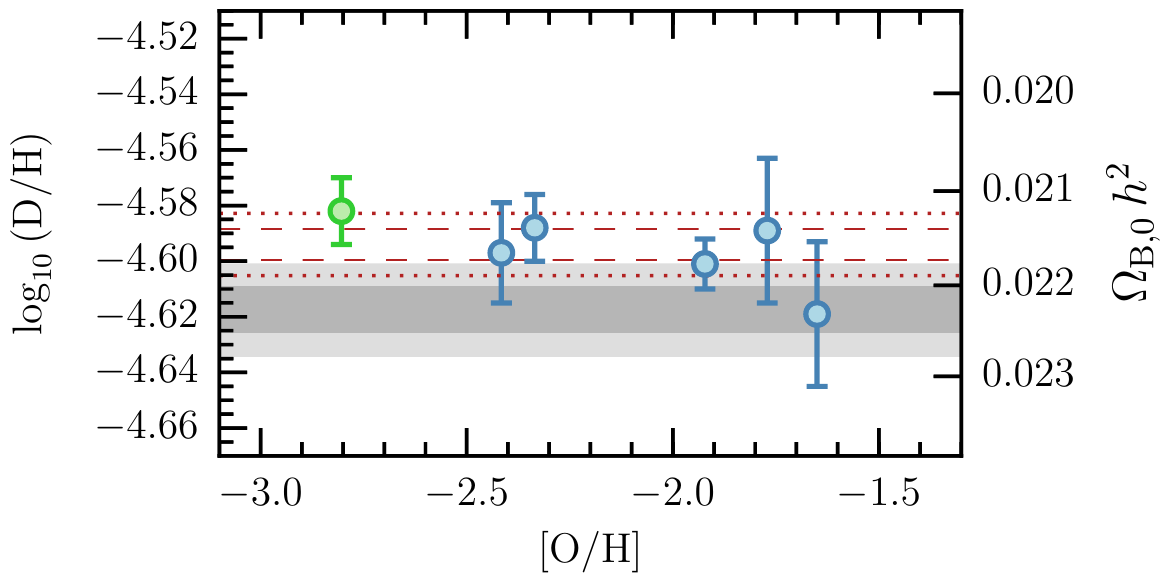}
 \hspace{6mm}\includegraphics[angle=0,width=85mm]{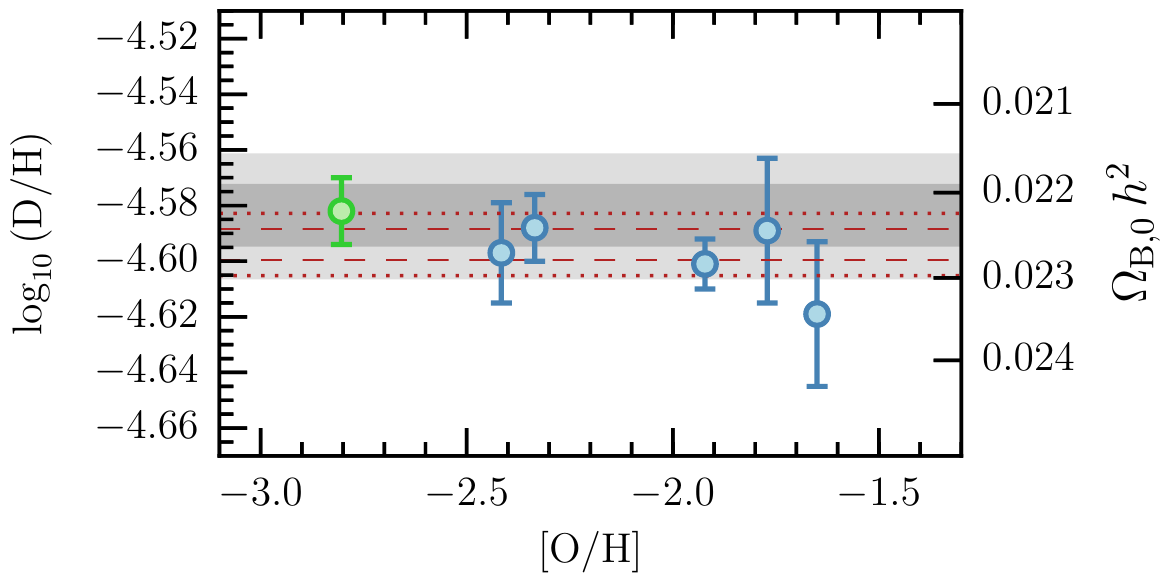}}\\
  \caption{
We plot the current sample of high quality primordial D/H abundance
measurements (symbols with error bars) as a function of the oxygen
abundance. The green symbol (with the lowest
value of [O/H]) corresponds to the new measurement reported
here, and the blue symbols are taken from \citet{Coo14}.
The red dashed and dotted
horizontal lines indicate the 68 and 95 per cent
confidence interval on the weighted mean value of the six
high precision D/H measures listed in Table~\ref{tab:dhmeasures}.
The right axes show the conversion between D/H and
\obhh\ for the Standard Model.
The conversion shown in the left panel uses the recent
theoretical determination of the $d(p, \gamma)^3$He
reaction rate (and its error) by \citet{Mar16},
while the right panel uses an empirical $d(p, \gamma)^3$He
rate and error based on the best available experimental data
(see \citet{NolBur00} and \citet{NolHol11} for a critical
assessment of the available experimental data).
In both panels, the gray horizontal band
shows the Standard Model D/H abundance based on our
BBN calculations (see text) and the universal baryon density
determined from the CMB temperature fluctuations \citep{Efs15}.
The dark and light shades of gray represent the 68 and 95 per cent
confidence bounds, respectively, including the uncertainty in the
conversion of \obhh\ to D/H (0.83 per cent for the left panel and
2.0 per cent for the right panel).
The Standard Model value displayed in the left panel is
0.02~dex lower in $\log_{10}$(D/H) than that shown in
Figure~5 of \citet{Coo14}. This shift is largely
due to the updated \textit{Planck} results \citep{Efs15}, and the
updated theoretical $d(p,\gamma)^{3}{\rm He}$ reaction rate
\citep{Mar16}.
  }
  \label{fig:measures}
\end{figure*}

In what follows, we only consider the six highest quality, and self-consistently
analyzed D/H abundance measurements; this sample includes the new
measurement that we report herein, and the sample of five measurements
previously analyzed by \citet{Coo14}. These measures are presented as
a function of [O/H] metallicity in Fig.~\ref{fig:measures}, and are listed in
Table~\ref{tab:dhmeasures}. For other recent D/H measures,
and a more complete list of literature measurements, see \cite{Rie15} and \citet{Bal15}.

A visual inspection of Figure~\ref{fig:measures} may suggest that
there is a mild evolution (decline) of D/H with metallicity, given that
the value deduced here for the lowest metallicity DLA is the highest
of the six high-precision measures. However, we caution that the
trend is not statistically significant, given the small size of the
current sample. Specifically, assuming a linear evolution of
the D/H abundance with metallicity, we find:
\begin{equation}
\label{eqn:linearevol}
\log_{10}\,({\rm D/H}) = (-4.583\pm0.010) - (2.8\pm2.0)\times10^{3}({\rm O/H})
\end{equation}
where ${\rm (O/H)}=10^{\rm [O/H]-3.31}\equiv N$(\OI)/$N$(\HI).
The $p$-value of a non-evolving D/H ratio (rather
than a linear evolution with O/H) is 0.15, indicating
that our null hypothesis (the D/H abundance is
constant over the metallicity range of our sample)
can only be rejected at the $1.4 \sigma$ significance
level.

It is intriguing that the tentative decline of D/H with
increasing metallicity is in the same sense as expected
from galactic chemical evolution. On the other hand,
published models of the astration of D
(see \citealt{Cyb15} for a list of references) do
not predict any significant evolution over the metallicity
range relevant here.
For example, the recent galactic chemical evolution models
of \citet{Wei16} entertain very minor corrections for astration
at the metallicities of the DLAs considered here
\citep[see also][]{Rom06,Dvo16}. Specifically the
D/H astration correction is estimated to be 0.33 per cent
and 0.023 per cent (+0.0015 and +0.0001 in the log) from
the least to the most metal-poor DLA listed in Table~\ref{tab:dhmeasures}.
These (systematic) upward corrections to D/H are
significantly smaller than the random errors
associated with the six measures of D/H.

For comparison, converting Equation~16 of \citet{Wei16} into the
form of our Equation~\ref{eqn:linearevol}, we estimate a slope of
$\approx-140$ for their fiducial model, which is a factor of $\sim20$
lower than the value estimated using the observational data
(see Equation~\ref{eqn:linearevol}). This suggests that astration
is not responsible for the mild evolution of D/H with metallicity
(if there is one at all over the range of O/H values of our sample).

Another possibility is that deuterium may be preferentially
depleted onto dust grains \citep{Jur82,Dra04,Dra06}. This
effect has been seen in the local interstellar medium of the
Milky Way \citep{Woo04,ProTriHow05,Lin06,EliProLop07,LalHebWal08,ProSteFie10}.
However, unlike the Milky Way, the DLAs that we investigate
here are very low metallicity ([Fe/H]~$<-2.0$); even the most
refractory elements in such DLAs exhibit negligible dust
depletions \citep{Pet97,Vla04,Ake05}, and very low
metallicity DLAs are not expected
to harbor a significant amount of dust
\citep[see][and references therein]{MurBer16}.
Ultimately, this issue will be clarified by extending the
number of precision measures of D/H over a wider
range of metallicity than covered by the present sample.

\begin{table*}
\begin{center}
    \caption{\textsc{precision d/h measures considered in this paper}}
    \hspace{-0.6cm}\begin{tabular}{@{}lccccc}
    \hline
   \multicolumn{1}{c}{QSO}
& \multicolumn{1}{c}{$z_{\rm em}$}
& \multicolumn{1}{c}{$z_{\rm abs}$}
& \multicolumn{1}{c}{log~$N$(\HI)/cm$^{-2}$}
& \multicolumn{1}{c}{[O/H]$^{\rm a}$}
& \multicolumn{1}{c}{log~\DI/\HI}\\
  \hline
HS\,0105$+$1619          &  $2.652$  &  $2.53651$  &  $19.426\pm0.006$  &  $-1.771\pm0.021$  &  $-4.589\pm0.026$  \\
Q0913$+$072                 &  $2.785$  &  $2.61829$  &  $20.312\pm0.008$  &  $-2.416\pm0.011$  &  $-4.597\pm0.018$  \\
SDSS~J1358$+$0349  &  $2.894$  &  $2.85305$  &  $20.524\pm0.006$   &  $-2.804\pm0.015$  &  $-4.582\pm0.012$  \\
SDSS~J1358$+$6522  &  $3.173$  &  $3.06726$  &  $20.495\pm0.008$   &  $-2.335\pm0.022$  &  $-4.588\pm0.012$  \\
SDSS~J1419$+$0829  &  $3.030$  &  $3.04973$  &  $20.392\pm0.003$   &  $-1.922\pm0.010$  &  $-4.601\pm0.009$  \\
SDSS~J1558$-$0031   &  $2.823$  &  $2.70242$  &  $20.75\pm0.03$        &  $-1.650\pm0.040$  &  $-4.619\pm0.026$  \\
  \hline
    \end{tabular}
    \label{tab:dhmeasures}

$^{\rm a}${We adopt the solar value log\,(O/H) + 12 = 8.69 \citep{Asp09}.}\\
\end{center}
\end{table*}

\subsection{Implications for Cosmology}

As discussed above, the six self-consistently analyzed D/H
abundance measurements that we consider here are statistically
consistent with being drawn from the same value. Hereafter,
we assume that all six measures provide a reliable estimate
of the primordial abundance of deuterium, $({\rm D/H})_{\rm P}$.
From the weighted mean of these independent values we
deduce our best estimate of the primordial deuterium abundance:
\begin{equation}
\label{eqn:dhp}
\log_{10}\,({\rm D/H})_{\rm P} = -4.5940\pm0.0056
\end{equation}
or, expressed as a linear quantity:
\begin{equation}
10^{5}\,({\rm D/H})_{\rm P} = 2.547\pm0.033.
\end{equation}

To compare our determination of $({\rm D/H})_{\rm P}$ with the latest \textit{Planck}
CMB results, we computed a series of detailed BBN calculations
that include the latest nuclear physics input.
Our simulation suite is identical to that described by
\citet{NolBur00}, but includes updates to:
(1) The neutron lifetime from \citet{PDG14};
(2) new experimental cross section measurements for $d(d,n)^{3}{\rm He}$, $d(d,p)^{3}{\rm H}$ \citep{Gre95,Leo06},
and $^{3}{\rm He}(\alpha,\gamma)^{7}{\rm Be}$ \citep{CybFieOli08,Ade11}; and
(3) new theoretical cross section calculations of $p(n,\gamma)d$ \citep{Rup00}
and $d(p,\gamma)^3\mathrm{He}$ \citep{Mar16}. For further details on all but
$d(p,\gamma)^3\mathrm{He}$, see \citet{NolHol11}.

The $d(p,\gamma)^3\mathrm{He}$
reaction rate can now be reliably computed with a precision of about 1 per cent,
compared with current laboratory measurements that have an uncertainty of $\gtrsim7$ per cent.
Our previous work used the $d(p,\gamma)^3\mathrm{He}$ reaction rate calculated by \citet{Mar05}.
Recently, \citet{Mar16} have published a revised calculation, which includes
a $\sim 2.5$ per cent relativistic correction that had previously been found
to be large in $d(n,\gamma)^3\mathrm{H}$. 
The new calculation also includes a quantitative error estimate that
is better than 1 per cent at most energies and incorporates wave
functions that have been extensively tested for accuracy.
We use the numerical uncertainty quoted by \citet{Mar16},
and do not use laboratory data to inform the theoretical rate
(see e.g. \citealt{Coc15}); at BBN energies, the laboratory data predominantly
consist of one experiment that has relatively low precision and is in moderate
conflict with the calculation. For comparison, we also consider how the
output nucleosynthesis is altered if we use the empirical
$d(p,\gamma)^3\mathrm{He}$ reaction rate instead of
the theoretical rate (see below).
Although we use the numerical uncertainty quoted by \citet{Mar16},
it should be pointed out that no quantitative estimate
exists for further uncertainties in construction of the
nucleon-nucleon potential and current operators, which could be of
similar size.  We have attempted to account for some of this with a
0.5 per cent correlated error on all points of the curve.

We now describe a summary of our BBN calculations, and
direct the reader to \citet{NolBur00} for further
details. First, the calculations are initialized with a Gaussian
random realization of each cross section measurement or (in the cases
of $p(n,\gamma)d$ and $d(p,\gamma)^3\mathrm{He}$) calculation.  The
distributions of point-to-point errors and of the (usually larger)
normalization errors shared by all points from a given experiment are
sampled independently.  Then a continuous, piecewise polynomial is fit
to the sampled cross sections. The thermal reaction rates at BBN
temperatures are calculated for each realization, using the sampled
and fitted cross sections. These rates are used as input into a BBN
code, along with a Gaussian random realization of the neutron
lifetime, and the output nucleosynthesis is stored.
At a given value of the expansion rate (parameterized by the
number of neutrino species, $N_{\nu}$\footnote{Our BBN model
includes the effect of incomplete neutrino decoupling, which
makes $N_\mathrm{eff} \neq 3$ at recombination for the
Standard Model, as a small additive correction to the $Y_\mathrm{P}$ yield.
BBN yields away from the Standard Model are computed by rescaling the
neutrino energy density during BBN by a factor $N_\nu/3$. We then assume that
the expansion rate at recombination is governed by an effective number of neutrino
species, $N_\mathrm{eff} = 3.046 N_\nu /3$ (\citealt{Man05}; see also, \citealt{Gro15}).
To the best of our knowledge, no detailed calculation of neutrino
weak decoupling has been published for expansion rates equivalent to
$N_\nu \neq 3$.}) and the density ratio
of baryons-to-photons ($\eta_{10}$, in units of $10^{10}$), we
perform 24\,000 Monte Carlo realizations, which was deemed to
provide smooth $2\sigma$ confidence contours as a function of
$\eta_{10}$ \citep[see][]{NolBur00}. This procedure provides a
thorough accounting of the current error budget for primordial
nucleosynthesis calculations.

We computed the resulting nucleosynthesis over the range
$1.8\le{N_{\nu}}\le4$ (in steps of $0.2$) and
$0.477\le\log_{10}\,\eta_{10}\le1.0$ (in steps of $\sim0.026$), and
interpolated this two-dimensional grid with a cubic spline.
Our interpolated grid of values is accurate to within $0.1$
per cent. For a given \neff\ and \obhh, the final
distribution of D/H values is Gaussian in shape, and
offers an uncertainty on (D/H)$_{\rm P}$ of $\lesssim1$ per cent
over the full parameter grid;
for the Standard Model, the uncertainty of the primordial
deuterium abundance is $\sim0.83$ per cent when using the
theoretical $d(p,\gamma)^3\mathrm{He}$ reaction rate. For convenience,
we also provide the following simple fitting formula that describes
how the D/H abundance depends on $\eta_{10}$ and \neff:
\begin{equation}
\label{eqn:dhconv}
10^{5}\,({\rm D/H})_{\rm P} = 2.47\,(1\pm0.01)\,(6/\eta_{\rm D})^{1.68}
\end{equation}
where
\begin{equation}
\eta_{\rm D} = \eta_{10} - 1.08\,(S-1)\,(1.1\,\eta_{10}-1)
\end{equation}
\begin{equation}
\label{eqn:dhconvb}
S = \Big(1+\frac{7\Delta N_{\nu}}{43}\Big)^{1/2}
\end{equation}
and $N_{\rm eff} = 3.046\,(1 + \Delta N_{\nu}/3)$. This functional form
is a slightly modified version of the form introduced by \citet{KneSte04},
and is accurate to within 0.4 per cent over the range 
$2.3\le{N_{\rm eff}}\le3.7$ and $5.4\le\eta_{10}\le6.6$.
The uncertainty quoted in Equation~\ref{eqn:dhconv}
includes both the 0.4 per cent uncertainty in the form of the
fitting function as well as the uncertainty in the BBN calculation.

To convert the baryon-to-photon ratio into a measurement of
the cosmic density of baryons, we use the conversion
$\eta_{10}=(273.78\pm0.18)\times\Omega_{\rm B,0}\,h^{2}$ \citep{Ste06},
which assumes a primordial helium mass fraction
$Y_{\rm P}=0.2471\pm0.0005$ (see Equation 43-44 from \citealt{LopTur99})
and a present day CMB photon temperature
$T_{\gamma,0}=2.72548\pm0.00057$ \citep{Fix09}.
Using the weighted mean value of the primordial
deuterium abundance (Equation~\ref{eqn:dhp}),
we estimate the cosmic density of baryons for the
Standard Model:
\begin{equation}
\label{eqn:obhhbbn}
100\,\Omega_{\rm B,0}\,h^{2}({\rm BBN}) = 2.156\pm0.017\pm0.011
\end{equation}
where the first error term includes the uncertainty in the
measurement and analysis, and the second error term
provides the uncertainty in the BBN calculations.
This level of precision is comparable to or somewhat
better than that achieved by the latest data release from
the \textit{Planck} team \citep{Efs15}.

The value of $\Omega_{\rm B,0}\,h^{2}({\rm BBN})$ reported here
(Equation~\ref{eqn:obhhbbn}) differs from the one reported by
\citet{Coo14} in two ways: (1) Our new measure of $\Omega_{\rm B,0}\,h^{2}({\rm BBN})$
is lower by 0.00046 (i.e. a $\sim2.1$ per cent change); and (2) the measurement
uncertainty is now the dominant term of the total error budget, whereas
the earlier estimate was dominated by the uncertainty in the BBN calculations.
The reduced uncertainty here results from using the \citet{Mar16}
$d(p,\gamma)^3\mathrm{He}$ cross section and its estimated
$\sim1$ per cent error.  Previously, we used the \citet{Mar05} calculation,
which lacked a quantitative error estimate.\footnote{Our previous estimate
of the $d(p,\gamma)^3\mathrm{He}$ cross section uncertainty was based on
experimental cross section measurements below the BBN energy range
(with an error of 7 per cent). Note that both the \citet{Mar05} and \citet{Mar16}
calculations agree closely with these low energy experimental data \citep{NolHol11}.}
The new calculation also reduces the D yield slightly through a
combination of a better electromagnetic current operator and more
careful attention to the wave function precision.\footnote{\citet{Mar16}
also present BBN calculations based on their
new cross sections, using the Parthenope code \citep{Pis08}.
At the Planck baryon
density, they now find $(\mathrm{D/H})_\mathrm{P} = 2.46\times
10^{-5}$ after a small change to their code (Marcucci 2016, private
communication).  Using either the \citet{Mar16} rate or the
\citet{Ade11} rate for $d(p,\gamma)^3\mathrm{He}$, there is
a consistent 2\%\ difference between their BBN code and ours.}
The \citet{Mar16} cross section calculation
results in a change to both the normalization and shape of the D/H abundance as a
function of $\eta_{10}$; for the Standard Model, the primordial D/H abundance is
shifted by $2.6$ per cent, and the uncertainty of this reaction rate is reduced by a factor of
$\sim4$ relative to that used by \citet{Coo14}.

The Standard Model value of the cosmic baryon density
obtained from our BBN analysis is somewhat lower than
that extracted from the temperature fluctuations of the CMB,
$100\,\Omega_{\rm B,0}\,h^{2}({\rm CMB})=2.226\pm0.023$
\citep[][see gray bands in Fig.~\ref{fig:measures}]{Efs15}\footnote{This
value of \obhh\ corresponds to the TT+lowP+lensing analysis
 (i.e. the second data column of Table~4 from \citealt{Efs15}).}.
This difference corresponds to a $2.3\sigma$ discrepancy
between BBN and the CMB for the Standard Model.
If we consider the \textit{Planck} fits that include high-$l$
polarization, the significance of the disagreement becomes $2.7\sigma$
(TT,TE,EE+lowP+lensing), or $3\sigma$ in combination with external
data (TT,TE,EE+lowP+lensing+ext).
We also note that the central value of \obhh\ derived from the
\textit{Planck} CMB is robust; the \textit{Planck} team consider a series of one
parameter extensions to the base $\Lambda$CDM model and
in all cases, the uncertainty on \obhh\ is inflated but
the central value remains unchanged.

\begin{figure}
  \centering
 {\includegraphics[angle=0,width=80mm]{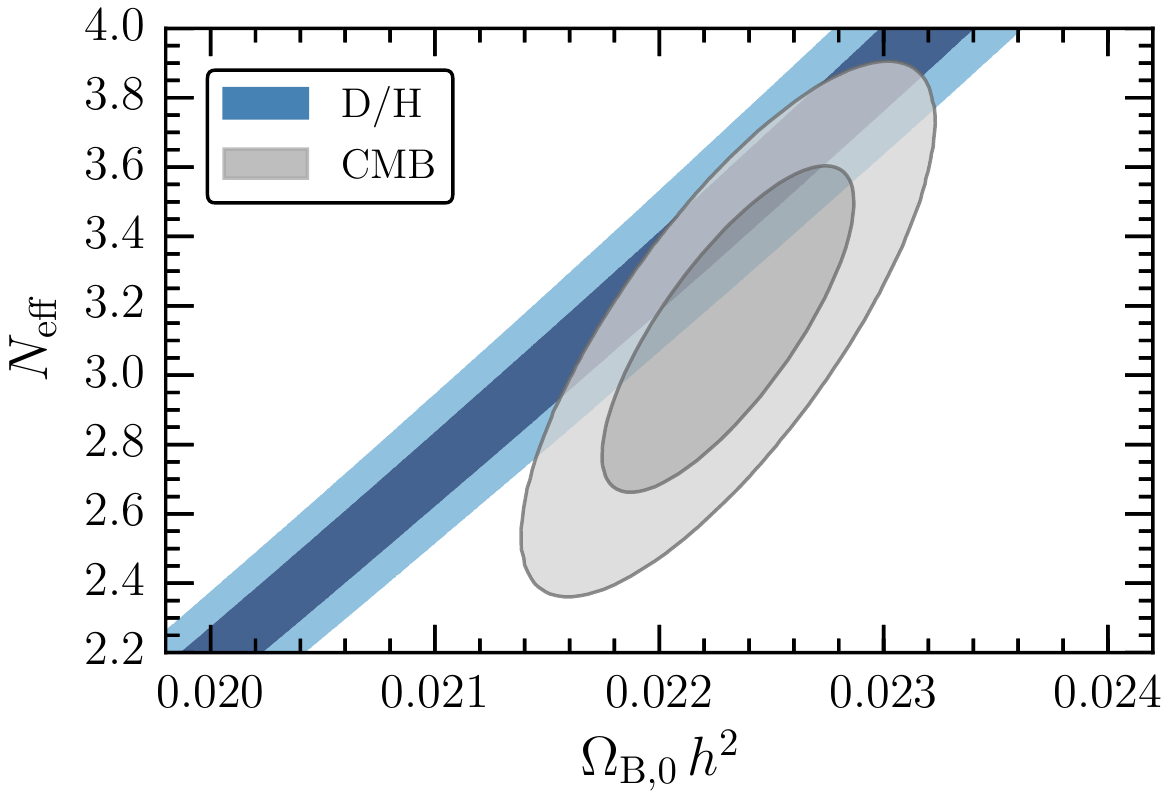}}\\
  \caption{
Comparing the expansion rate (parameterized by \neff) and the
cosmic density of baryons (\obhh) from BBN (blue contours) and
CMB (gray contours). The dark and light shades
illustrate the 68\%\ and 95\%\ confidence contours, respectively.
  }
  \label{fig:neffobhh}
\end{figure}

By considering a deviation in the Standard Model expansion
rate of the Universe, as parameterized by \neff, the significance
of the disagreement between CMB and BBN is reduced to the
$1.5\sigma$ level.\footnote{The disagreement becomes more
significant ($2.4\sigma$) if we consider the
\textit{Planck} TT,TE,EE+lowP analysis.}
This comparison is shown in Fig.~\ref{fig:neffobhh}
for the \textit{Planck} TT+lowP analysis \citep[for similar comparisons
between CMB and BBN constraints, see][]{Efs14,Efs15,Coo14,NolSte15,Cyb15}.
If we assume that \neff\ and \obhh\ do not change from BBN to recombination,
the combined confidence bounds on the baryon density and the
effective number of neutrino families are (95 per cent confidence limits):
\begin{eqnarray}
100\,\Omega_{\rm B,0}\,h^{2} &=& 2.235\pm0.071\\
N_{\rm eff} &=& 3.44\pm0.45.
\end{eqnarray}

The aforementioned disagreement between the CMB and BBN has emerged
as a result of the improved reaction
rate calculation reported recently by \citet{Mar16}.
To show the change introduced by this new rate, we have repeated our BBN
calculations using an empirically derived $d(p,\gamma)^3\mathrm{He}$
rate, in place of the theoretical rate. We use
all published data that are credible as absolute cross sections
\citep{Gri62,Sch97,Ma97,Cas02},\footnote{
Of these, only \citet{Ma97} probe the key energy range of late-BBN
deuterium burning; see \citet{NolBur00} and \citet{NolHol11} for
further details.}
and generate Monte Carlo realizations of these experimental data, as described above.
Our BBN calculations, combined with our measurement of the
primordial D/H abundance (Equation~\ref{eqn:dhp}), return
a Standard Model value of the cosmic baryon density:
\begin{equation}
\label{eqn:obhhbbne}
100\,\Omega_{\rm B,0}\,h^{2}({\rm BBN}) = 2.260\pm0.018\pm0.029
\end{equation}
which is in somewhat better agreement with the \citet{Efs15} value,
albeit with a much larger nuclear error (i.e. the second
error term in Equation~\ref{eqn:obhhbbne}).\footnote{The
data-driven Monte Carlo procedure that we use here has greater
freedom to match $S$-factor data than the widely used quadratic fit of
\citet{Ade11}, resulting in a somewhat lower $d(p,\gamma)^3\mathrm{He}$
rate.  Adopting the \citet{Ade11} $S$-factor curve would
change Equation~\ref{eqn:obhhbbne} to
$100\,\Omega_{\rm B,0}\,h^{2}({\rm BBN}) = 2.225\pm0.018\pm0.033$.}

In the right panel of Fig.~\ref{fig:measures}, we compare our D/H
measurements to the Standard Model deuterium abundance based
on the \citet{Efs15} baryon density and our calculations that use
the empirical $d(p,\gamma)^3\mathrm{He}$ rate.
Using the empirical rate
shifts the Standard Model value of the primordial D/H abundance
upwards by $\sim8$ per cent, and inflates the corresponding
uncertainty by a factor of $\sim1.5$.

At present, it is difficult to tell how seriously to interpret the
discrepancy between BBN and the CMB. Doubling the estimated
nuclear error in Equation~\ref{eqn:obhhbbn} still leaves us with a
$2\sigma$ disagreement (assuming $N_\mathrm{eff} = 3.046)$.
This doubling would require a
$\sim4$ per cent error on $d(p,\gamma)^3\mathrm{He}$, which
seems a large overestimate relative to the $\sim1$ per cent
errors quoted by \citet{Mar16}\footnote{Similarly, other relevant
reaction rates, such as $d+d$,
have been measured in the laboratory with high precision
and are unlikely to contribute significantly to the error
budget.}.
Alternatively, the CMB and BBN would agree exactly if the
\citet{Mar16} rate was scaled downwards by $\sim10$ per cent
\citep[see e.g.][]{DiV14,Efs15}; however, a
significant change to the rate normalization is unlikely, given
the accuracy with which rates can now be calculated for a
three-body system \citep{Kie08}.
It is helpful that the lack of empirical information on
$d(p,\gamma)^3\mathrm{He}$ at BBN energies is currently being
addressed by the LUNA collaboration \citep{Gus14}.
However, if they achieve high precision, their result seems
unlikely to fit well with both cosmology and nuclear theory
simultaneously.

\section{Summary and Conclusions}
\label{sec:conc}

Several probes of cosmology have now pinned down the
content of the Universe with exquisite detail. In this paper,
we build on our previous work to obtain precise
measurements of the primordial deuterium abundance
by presenting high quality spectra of a DLA at $z_{\rm abs}=2.852054$
towards the quasar J1358$+$0349, taken with both the
UVES and HIRES instruments. Our primary
conclusions are as follows:\\

\noindent ~~(i) The absorption system reported here is
the most metal-poor DLA currently known, with an average
oxygen abundance [O/H]~$= -2.804\pm0.015$.
Furthermore, in one of the absorption components, we
estimate [O/H]~$= -3.07\pm0.03$. This environment is
therefore ideally suited to estimate the primordial
abundance of deuterium.
On the other hand, we have found an unusual amount
of unrelated absorption that contaminates many of the weak,
high order, \DI\ absorption lines. Consequently, the accuracy
in the determination of the D/H ratio achieved for this system
is not as high as the best cases reported by
\citet[][J1419$+$0829]{PetCoo12a} and \citet[][J1358$+$6522]{Coo14},
see Table~\ref{tab:dhmeasures}.

\smallskip

\noindent ~~(ii) Using an identical analysis strategy to that described
in \citet{Coo14}, we measure a D/H abundance of
$\log_{10}\,({\rm D\,\textsc{i}/H\,\textsc{i}}) = -4.582\pm0.012$
for this near-pristine DLA. We estimate that this abundance
ratio should be adjusted
by $(-4.9\pm1.0)\times10^{-4}$~dex to account for \DII\ charge
transfer recombination with \HI. This ionization correction
is a factor of $\sim25$ less than the D/H measurement
precision of this system, and confirms that 
${\rm D\,\textsc{i}/H\,\textsc{i}}\cong{\rm D/H}$ in DLAs.

\smallskip

\noindent ~~(iii) On the basis of six high precision and self-consistently
analyzed D/H abundance measurements, we report tentative evidence
for a decrease of the D/H abundance with increasing metallicity. If confirmed,
this modest decrease of the D/H ratio could provide an important opportunity
to study the chemical evolution of deuterium in near-pristine environments.

\smallskip

\noindent ~~(iv) A weighted mean of these six independent D/H measures
leads to our best estimate of the primordial D/H abundance,
$\log_{10}\,({\rm D/H})_{\rm P} = -4.5940\pm0.0056$. We combine
this new determination of (D/H)$_{\rm P}$ with a suite of detailed
Monte Carlo BBN calculations. These calculations include updates
to several key cross sections, and propagate the uncertainties of
the experimental and theoretical reaction rates.
We deduce a value of the cosmic baryon
density $100\,\Omega_{\rm B,0}\,h^{2}({\rm BBN}) = 2.156\pm0.017\pm0.011$,
where the first error term represents the D/H measurement uncertainty
and the second error term includes the uncertainty of the BBN calculations.

\smallskip

\noindent ~~(v) The above estimate of \obhh(BBN) is comparable in
precision to the recent determination of \obhh\ from the
CMB temperature fluctuations
recorded by the \textit{Planck} satellite. However, using the best
available BBN reaction rates, we find a $2.3\sigma$ difference
between \obhh(BBN) and \obhh(CMB), assuming the Standard
Model value for the effective number of neutrino species,
$N_{\rm eff}=3.046$. Allowing \neff\ to vary, the disagreement
between BBN and the CMB can be reduced to the $1.5\sigma$
significance level, resulting in a bound on the effective number
of neutrino families, $N_{\rm eff} = 3.44\pm0.45$.

\smallskip

\noindent ~~(vi) By replacing the theoretical $d(p,\gamma)^{3}{\rm He}$
cross section with the current best empirical estimate, we derive a baryon
density $100\,\Omega_{\rm B,0}\,h^{2}({\rm BBN}) = 2.260\pm0.034$, which
agrees with the \textit{Planck} baryon density for the Standard Model. However,
this agreement is partly due to the larger error estimate for the nuclear data.
Forthcoming experimental measurements of the crucial $d(p,\gamma)^{3}{\rm He}$
reaction rate by the LUNA collaboration will provide important
additional information regarding this discrepancy, since the empirical rate
currently rests mainly on a single experiment, and absolute cross
sections often turn out in hindsight to have underestimated errors.
The theory of few-body nuclear systems is now precise enough that a
resolution in favor of the current empirical rate would present a
serious problem for nuclear physics.

\smallskip

Our study highlights the importance of expanding the present
small statistics of high precision D/H measurements, in
combination with new efforts to achieve high precision in
the nuclear inputs to BBN. We believe that
precise measurements of the primordial D/H abundance
should be considered an important goal for the future
generation of echelle spectrographs on large telescopes,
optimized for wavelengths down to the atmospheric cutoff.
This point is discussed further in Appendix~\ref{app:future}.

\section*{Acknowledgements}
We are grateful to the staff astronomers at the VLT \&\ Keck  
Observatories for their assistance with the observations, and
to Jason X. Prochaska and Michael Murphy for providing some
of the software that was used to reduce the data. We thank
Gary Steigman for interesting discussions, and useful comments
on an early draft.
We also thank an anonymous referee who provided helpful
suggestions that improved the presentation of this work.
RJC is currently supported by NASA through
Hubble Fellowship grant HST-HF-51338.001-A, awarded by the
Space Telescope Science Institute, which is operated by the
Association of Universities for Research in Astronomy, Inc.,
for NASA, under contract NAS5- 26555. RAJ acknowledges
support from an NSF Astronomy and Astrophysics Postdoctoral
Fellowship under award AST-1102683. We thank the Hawaiian
people for the opportunity to observe from Mauna Kea;
without their hospitality, this work would not have been possible.
RJC thanks JBC for his impeccable timing and invaluable insight.


\begin{appendix}

\section{A. Measuring D/H with future facilities}
\label{app:future}

The six, high quality D/H measurements considered in this
work were all observed with 8--10\,m class telescopes equipped
with efficient echelle spectrographs. In this Appendix, we
estimate how the D/H sample size scales with telescope
aperture and UV wavelength coverage. This calculation
provides an indicative number of D/H systems that
will be accessible to the next generation of 30--40\,m
class telescopes.

Starting with the \citet{Ros13} redshift dependent quasar
luminosity function, we calculate the redshift distribution of
quasars brighter than those accessible with the 8--10\,m class
telescopes (i.e. those with apparent magnitude $m_{\rm ref}\lesssim19$).
The magnitude limit that we have chosen corresponds to the limiting
magnitude with which data of sufficient signal-to-noise ratio
(i.e. ${\rm S/N}\gtrsim10$ at the bluest wavelengths) can be
acquired within 1 night of 8--10\,m telescope time. We integrate the
quasar luminosity function over the redshift interval
$z_{\rm lim} < z < 3.5$ where $2.7 < z_{\rm lim} < 3.5$;
our chosen $z_{\rm lim}$ range is based on the detectability
of the important, high order weak \DI\ lines near 915\,\AA.
At redshifts $z_{\rm lim}\lesssim2.7$, the Earth's atmosphere
significantly absorbs the bluest light of an $m_{\rm ref}\simeq19$
background quasar (i.e. at wavelengths less than
$915\,\AA\times(1+2.7)\simeq3400$\,\AA), making it
impossible to reach the required S/N near the weak \DI\
absorption lines in a reasonable amount of time.
At the other limit, when $z_{\rm lim}\gtrsim3.5$,
the \Lya\ forest increasingly contaminates the high
order \DI\ lines (discussed further below); note that
our calculation is largely insensitive to the chosen
upper limit on $z_{\rm lim}$.

The calculated redshift distribution of quasars brighter
than an apparent magnitude $m_{\rm ref}=19$ is then
normalized to one. We use this normalization factor as
our reference value to scale the remaining results of our
calculation. This factor allows us to estimate the number
of D/H measurements that can be made with future
facilities for every one system that can be observed
with the current 8--10\,m telescopes.

We now estimate the magnitude limit of the quasars that
are accessible to the three, currently planned,
next generation telescope facilities:
(1) The European Extremely Large Telescope (E-ELT),
with a collecting area of 978\,m$^{2}$,
(2) The Thirty Meter Telescope (TMT),
with a collecting area of 655\,m$^{2}$, and
(3) The Giant Magellan Telescope (GMT),
with a collecting area of 368\,m$^{2}$.
Assuming all other factors to be equal, we scale
the apparent magnitude limit of an 8--10\,m
class telescope by its collecting area:
\begin{equation}
\label{eqn:apgain}
m_{\rm next}=m_{\rm ref} + 2.5\,\log_{10}({\cal A}/{\cal A}_{\rm ref})
\end{equation}
where ${\cal A}$ is the collecting area of a next generation
telescope, and ${\cal A}_{\rm ref}=76\,{\rm m}^{2}$
is the collecting area of the Keck telescope.
For the future telescopes, the typical magnitude
limit is in the range $m_{\rm next}\simeq21 - 22$.

This magnitude limit is comparable to the U-band brightness
of a dark night sky. In order to meet our limiting magnitude
criteria stated above (i.e. to acquire data of ${\rm S/N}\gtrsim10$
near the high order \DI\ lines), we must scale $m_{\rm next}$ (which
does not include the sky background) to the limiting magnitude,
$m_{\rm lim}$ (which includes the sky background). To estimate
this conversion, consider the S/N equation with and without a sky
background term, and demand that both cases yield the same
S/N ratio:
\begin{equation}
\label{eqn:equatesn}
\sqrt{F_{\rm next}} = \frac{F_{\rm lim}}{\sqrt{F_{\rm lim}+F_{\rm sky}}}
\end{equation}
where the desired limiting magnitude
$m_{\rm lim} = m_{\rm next} - 2.5\,\log_{10}(F_{\rm lim}/F_{\rm next})$,
and $F_{\rm sky}$ is the sky contribution to the total flux. Solving
Equation~\ref{eqn:equatesn} for $F_{\rm lim}/F_{\rm next}$ gives:
\begin{equation}
\label{eqn:flimnext}
F_{\rm lim}/F_{\rm next} = 0.5 + 0.5\sqrt{1+ 4\times10^{(m_{\rm next}-m_{\rm sky})/2.5}}
\end{equation}
where $m_{\rm sky}=22.35~{\rm arcsec}^{-2}$ is the typical U-band brightness
of a dark night sky \citep[e.g.][]{Pat08}.

Assuming a Gaussian seeing profile of 0.5~arcsec FWHM, and
a 0.7~arcsec entrance slit to the spectrograph, a projected slit
of $1\times0.7~{\rm arcsec}^{2}$ contains $\sim88$~per~cent
of the incident quasar flux; this reduction of 12~per~cent is
equivalent to increasing $m_{\rm next}$ in the exponent of
Equation~\ref{eqn:flimnext} by $+0.14~{\rm mag}$.
Similarly, the U-band sky brightness within the projected slit is
30 per cent lower than the value quoted above for a
$1\times1~{\rm arcsec}^{2}$ aperture,
corresponding to a U-band sky magnitude
$m_{\rm sky}=22.74$.

The estimated limiting apparent magnitudes for the three future facilities are
$m_{\rm lim}$(E-ELT,~TMT,~GMT)~$\simeq$~21.5, 21.1, 20.6.
We then integrate the \citet{Ros13} quasar luminosity function
over an apparent magnitude range brighter than the above
limits, and scale the results to the normalizing factor derived earlier
for the 8--10\,m class telescopes.

We now account for the \emph{relative} \Lya\ forest
contamination suffered by quasars over the redshift range
that is considered here ($2.7 < z < 3.5$); quasars at higher
redshift are more likely to have \Lya\ forest absorption
that may contaminate the high order \DI\ absorption lines.
To assess the relative contamination, we first need to
estimate the number of pixels that are uncontaminated
by \Lya\ forest absorption over the wavelength range of
the weak \DI\ absorption lines ($\approx915-930$~\AA\ rest frame),
for a DLA that has a redshift $z_{\rm abs} > z_{\rm em} - 0.2$
(where $z_{\rm em}$ is the redshift of the quasar); DLAs with
a redshift close to that of the quasar are the most suitable for
high precision measures of D/H, since these DLAs will exhibit
a cleaner \Lya\ absorption line profile, and their \DI\ lines are
less likely to be contaminated by \Lya\ forest absorption.
We note that the condition $z_{\rm abs} > z_{\rm em} - 0.2$
is satisfied by all six high precision measures considered in
the present work (see Table~\ref{tab:dhmeasures}).

We have estimated the fraction of uncontaminated pixels
using the Keck Observatory Database
of Ionized Absorption towards Quasars (KODIAQ)
sample \citep{Ome15}.
The KODIAQ database consists of 170 high quality, fully
reduced, continuum normalized echelle spectra of quasars,
including 93 quasars in the redshift range $2.7 < z < 3.5$.
Each of these 93 quasars was visually inspected to identify
the sightlines that \emph{do not} contain a DLA with a redshift
greater than $z_{\rm em}-0.2$, since we want to estimate the
severity of \Lya\ forest contamination in the absence of a DLA.
Our final sample consists of 49 quasars, which we split
into two redshift bins containing
29 quasars with $2.7<z_{\rm em}<3.1$ ($\langle z_{\rm em}\rangle = 2.86$) and
20 quasars with $3.1<z_{\rm em}<3.5$ ($\langle z_{\rm em}\rangle = 3.30$).
We then calculate the fraction of pixels that exceed a
normalized flux of 0.9 over the observed wavelength range
$(1+z_{\rm em}-0.2)\times915 < \lambda < (1+z_{\rm em})\times930$,
where the interval 915\,\AA--930\,\AA\ includes the rest frame wavelengths
of the weakest \DI\ absorption lines.
These pixels are deemed to be free of contaminating absorption.
For the low redshift subsample we estimate a fraction of
uncontaminated pixels ${\cal F}(>0.9)=0.31$; for the
high redshift subsample, ${\cal F}(>0.9)=0.21$.

\begin{figure}
  \centering
 {\includegraphics[angle=0,width=140mm]{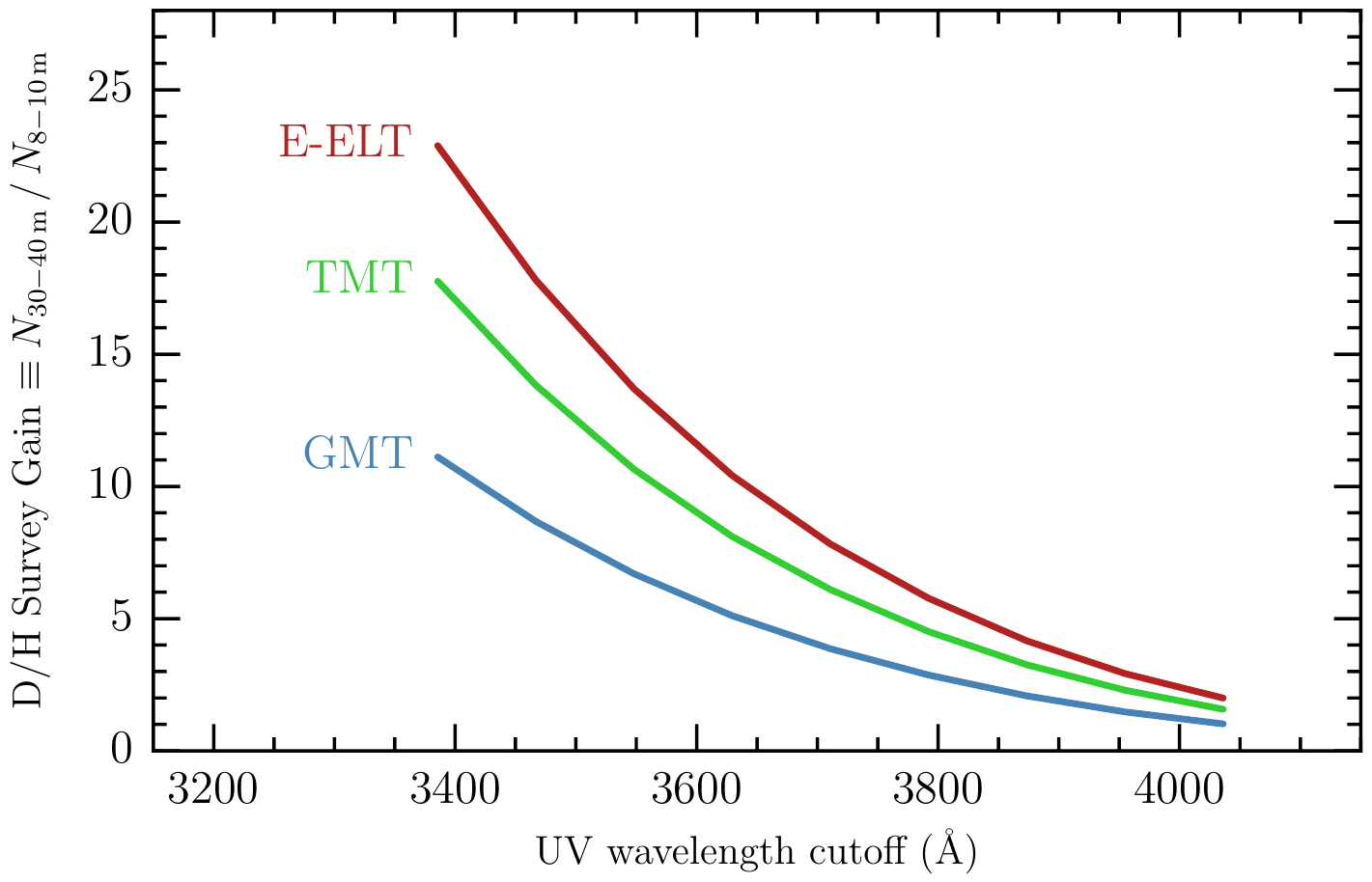}}\\
  \caption{
Each curve shows the number of D/H measurements
that can be made with future telescope facilities equipped
with an echelle spectrograph, relative to the number that
can be made with current facilities, as a function of the
`UV wavelength cutoff'.
The UV wavelength cutoff is defined as the observed
wavelength below which the redshifted high order
\DI\ lines ($\sim915$~\AA\ rest frame) cannot be
observed at S/N~$\gtrsim10$ within one night of
telescope time. The currently planned 30--40\,m
class telescopes include:
(1) The European Extremely Large Telescope (E-ELT; red, curve),
(2) the Thirty Meter Telescope (TMT; green curve) and
(3) the Giant Magellan Telescope (GMT; blue curve).
If the future facilities have sensitivity down to 3400\,\AA,
the current D/H sample size will be enlarged by at least
an order of magnitude.
  }
  \label{fig:future}
\end{figure}

The number of unblended \DI\ lines, $N$, is a
binomially distributed random variable. To obtain
a confident measure of the \DI\ column density, we
require that at least two of the five weakest \DI\
absorption lines are unblended, yielding the
probability ${\rm Pr}(N\ge2)=1-{\rm Pr}(N=0)-{\rm Pr}(N=1)$.
From this exercise, we estimate that
${\rm Pr}(N\ge2)_{z=2.86}/{\rm Pr}(N\ge2)_{z=3.30}=1.67$
(if we instead only require 1 \DI\ line to be unblended, the
relative probability is $1.20$).
Therefore, a quasar at $z=2.86$ is roughly 67 per cent
more likely to have at least two clean \DI\ lines, than a
quasar at $z=3.30$. To account for the increased relative
\Lya\ forest contamination at high redshift, we scale the
redshift distribution of quasars by the function
\begin{equation}
f(z)=\bigg(1+0.67\frac{3.30-z}{3.30-2.86}\bigg)~\bigg/~\bigg(1+0.67\frac{3.30-2.70}{3.30-2.86}\bigg)
\end{equation}
where the redshift, $z$, is related to the UV cutoff wavelength
by the equation $\lambda_{\rm cut}=915\times(1+z)$.
The result of the above calculation is shown in
Figure~\ref{fig:future}, where each curve illustrates the
number of D/H systems that will be accessible to each facility,
relative to the number that are accessible to current facilities,
as a function of the UV cutoff. For example, if 10 high
precision D/H measures can be made using 8--10\,m
telescopes, the 30--40\,m class telescopes could deliver
more than 200 high precision D/H measurements, provided
that the aperture gain (Equation~\ref{eqn:apgain}) is
maintained down to at least 3400\,\AA.
Similarly, if future D/H surveys are restricted to quasars
brighter than an apparent magnitude of 20.5 (equivalent
to the GMT curve in Figure~\ref{fig:future}), the current
statistics will be improved by over an order of magnitude.

As D/H measures are pushed towards higher precision,
it will become increasingly important to observe a large
sample of DLAs with diverse properties. This will allow
us to better understand potentially hidden systematics,
for example, due to ionization or chemical evolution.
Figure~\ref{fig:future} highlights the necessity for
efficient, UV-sensitive, high resolution spectrographs on
future 30-40\,m telescopes. If these capabilities can be
realized, it will become possible to significantly further
our measurements of D/H at high redshift, test for
departures from the Standard Model, and explore the
chemical evolution of galaxies via the astration of D.

\end{appendix}


\end{document}